# VSC-HVDC 系统的虚假数据注入攻击研究

韩通 [1]，陈艳波 [1]，马进 [2]

（1．华北电力大学电气与电子工程学院，新能源电力系统国家重点实验室，北京市 102206；2．School of Electric and Information Engineering，University of Sydney，NSW 2006，澳大利亚）

## Research on False Data Injection Attacks in VSC-HVDC Systems

HAN Tong[1], CHEN Yan-bo[1], MA Jin[2]

(1. State Key Laboratory of Alternate Electrical Power System with Renewable Energy Sources, North China Electric Power University, Beijing 102206, China; 2. School of Electric and Information Engineering, University of Sydney, NSW 2006, Australia)

**ABSTRACT:** The false data injection (FDI) attack is a crucial form of cyber-physical security problems facing cyber-physical power systems. However, there is no research revealing the problem of FDI attacks facing voltage source converter based high voltage direct current transmission (VSC-HVDC) systems. Firstly, the general form of the model of FDI attack strategies is proposed and the essence of the problem of FDI attack strategies is further analyzed. Moreover, the model of FDI attack strategies aiming at disrupting the operation security of converter stations in VSC-HVDC systems is proposed and its solving algorithm is then presented. And finally, the modified IEEE-14 bus system is utilized to reveal the problem of FDI attacks facing VSC-HVDC systems, demonstrating that attackers are capable of disrupting the operation security of converter stations in VSC-HVDC systems by FDI attacks.

**摘要：**虚假数据注入（False Data Injection，FDI）攻击作为电力信息-物理系统面临的信息物理安全问题的一种重要形式，受到了国内外研究人员日益广泛的关注。但目前尚无研究揭示基于电压源型换流器的高压直流输电（Voltage Source Converter based High Voltage Direct Current Transmission，VSC-HVDC）系统所面临的 FDI 攻击问题。本文首先提出了电力系统 FDI 攻击策略模型的一般形式，并对 FDI 攻击策略问题的实质进行了分析。进一步提出了以破坏 VSC-HVDC 系统换流器运行安全为目标的 FDI 攻击策略模型及求解算法。最后以修改的 IEEE-14 节点系统为例展示了 VSC-HVDC 系统面临的 FDI 攻击问题，结果表明攻击者可通过实施 FDI 攻击破坏 VSC-HVDC 系统的安全运行。

## 1 引言

随着信息技术的大量渗入，现代电力系统已经发展成为由电力及其监控系统构成的复杂信息-物理系统（Cyber-Physical System，CPS）[1-2]。由于信息系统与物理系统的高度耦合，信息系统遭受的故障或网络攻击不仅会影响其自身的正常运行，更有可能传导至物理系统从而威胁电力系统的安全、稳定和经济运行。这类新的安全问题被称为电力系统信息物理安全问题[3]。近几年国外发生的多起针对实际电网的网络攻击事件引起了国内外对电力 CPS 安全问题的广泛关注[4]。

电力系统借助数据采集与监控（Supervisory Control and Data Acquisition，SCADA）系统采集电网的实时数据并传输给电网调度中心的能量管理系统（Energy Management System，EMS），EMS 中的状态估计（State Estimation, SE）对含有误差和不良数据的量测生数据进行处理得到可信的系统状态变量的估计值，这一过程也称数据滤波，EMS 中的高级应用（如故障分析、自动发电控制、最优潮流等）进一步根据 SE 的结果实现对电力系统的闭环或半闭环控制。研究表明，当攻击者在掌握一定的系统信息时，可通过攻击 SCADA 系统中的量测仪表或通信链路，有目的地篡改 SCADA 系统传输给 EMS 的量测数据，从而影响 SE 的结果，进而

**基金项目：**国家自然科学基金资助项目(51407069)；中央高校基本科研业务费专项资金资助项目(2016YQ02)。

Project Supported by National Natural Science Foundation of China(NSFC)( 51407069); Fundamental Research Funds for the Central Universities(2016YQ02).

对 EMS 中的其他高级应用产生影响，并最终影响到整个电力 CPS 的安全、稳定和经济运行。这种通过篡改量测数据进而影响 SE 结果的网络攻击一般称之为虚假数据注入（False Data Injection，FDI）攻击，它是电力 CPS 所面临的信息物理安全问题的一种重要形式[5-6]。

自文献[7]首次揭示了电力 CPS 所面临的 FDI 攻击问题以来，国内外学者对其进行了大量研究。为简化分析，早期的文献大多基于直流 SE 模型来研究 FDI 攻击策略、FDI 攻击对电网的影响以及针对 FDI 攻击的应对策略[7-11]。但电力系统的非线性对于基于直流 SE 模型构建的 FDI 攻击策略具有一定的鲁棒性[12]，且实际电力系统中应用的 SE 模型一般为交流 SE 模型，针对直流 SE 构建的攻击策略未必有效。自文献[13]起，相关学者开始基于交流 SE 模型对 FDI 攻击进行研究。文献[13]从图论的角度研究了以攻击成本最小（最小化被攻击量测量数目）为目标的 FDI 攻击策略，但该 FDI 攻击策略只能对某一个电气量产生影响；文献[14]研究了在攻击者只掌握部分电网信息时，如何构建 FDI 攻击策略以对局部区域的负荷分布产生影响；文献[15]在文献[14]的基础上研究了最优攻击区域的确定方法，以使攻击者实施 FDI 攻击时所需掌握的电网信息最少；文献[16]研究了在攻击者无法掌握系统状态变量估计值情况下的 FDI 攻击策略。需要指出的是，已有文献均是对交流输电系统 FDI 攻击问题的研究，目前尚无研究揭示直流输电系统所面临的 FDI 攻击问题。直流输电系统的 FDI 攻击问题具有以下特点：（1）直流输电系统无对应的直流 SE 模型，且其状态变量、量测方程和等式约束亦有别于交流 SE 模型，因此无法构建基于直流 SE 的 FDI 攻击模型和策略，（2）基于交流 SE 的 FDI 攻击模型和策略也无法直接应用于直流输电系统。

近些年，基于电压源型换流器的高压直流输电（Voltage Source Converter based High Voltage Direct Current Transmission，VSC-HVDC）技术在电网中得到了广泛的应用，其在提高系统稳定性、增大输电能力以及促进新能源消纳等诸多方面发挥着重要作用。如果攻击者对 VSC-HVDC 系统实施 FDI 攻击，必将会对 VSC-HVDC 系统自身甚至整个交直流系统的安全稳定运行造成极大影响。研究 VSC-HVDC 系统的 FDI 攻击问题成为当务之急。

本文对 VSC-HVDC 系统所面临的 FDI 攻击问题进行研究。首先提出了电力系统 FDI 攻击策略模型的一般形式，并对 FDI 攻击策略问题的实质进行了分析；进一步提出以破坏 VSC-HVDC 系统换流器运行安全为目标的 FDI 攻击模型并给出了相应的求解算法；最后以修改的 IEEE-14 节点系统为例展示了 VSC-HVDC 系统面临的 FDI 攻击问题。

## 1 预备知识

### 1.1 含 VSC-HVDC 交直流系统的 SE 方法

含 VSC-HVDC 交直流系统大多采用加权最小二乘(Weight Least Squares, WLS)状态估计方法,其模型在交流系统 SE 模型的基础上增加了与 VSC-HVDC 系统相关的状态变量、量测方程及等式约束[17]。以下首先对 WLS 进行简要介绍，然后给出含 VSC-HVDC 交直流系统的量测方程。

一般地，SE 的量测方程和约束如式(1)所示:

$$\begin{cases} \boldsymbol{z}=\boldsymbol{h}(\boldsymbol{x})+\boldsymbol{e}+\boldsymbol{b} \\ \boldsymbol{c}(\boldsymbol{x})=\boldsymbol{0} \end{cases} \tag{1}$$

式中，$\boldsymbol{z}^T=[z_1, z_2, \dots, z_m]$ 和 $\boldsymbol{x}^T=[x_1, x_2, \dots, x_n]$ 分别表示量测向量和状态向量；$\boldsymbol{e}^T=[e_1, e_2, \dots, e_m]$ 表示量测误差向量，且满足 $e_i \sim N(0, \sigma_i^2)$，$i=1, \dots, m$；$\boldsymbol{b}^T=[b_1, b_2, \dots, b_m]$ 表示不良数据向量（相对值）；$\boldsymbol{h}^T=[h_1(\boldsymbol{x}), h_2(\boldsymbol{x}), \dots, h_m(\boldsymbol{x})]$，$h_i(\boldsymbol{x})$ 表示与 $z_i$ 对应的量测真值；$\boldsymbol{z}=\boldsymbol{h}(\boldsymbol{x})+\boldsymbol{e}+\boldsymbol{b}$ 表示量测方程；$\boldsymbol{c}(\boldsymbol{x})=\boldsymbol{0}$ 为等式约束。另外，本文涉及到等式约束处均将等式约束视作高精度的虚拟量测进行求解。WLS 通过求解下面的优化模型得到系统状态向量的估计值 $\hat{\boldsymbol{x}}$，即:

$$\hat{\boldsymbol{x}}=\arg\min_{\boldsymbol{x}} \left[\boldsymbol{z}-\boldsymbol{h}(\boldsymbol{x})\right]^T \boldsymbol{R}^{-1}\left[\boldsymbol{z}-\boldsymbol{h}(\boldsymbol{x})\right] \tag{2}$$

式中，$\boldsymbol{R}$ 为量测向量 $\boldsymbol{z}$ 的误差方差阵。

WLS 在实际应用中须配置有不良数据检测与辨识环节，最为常见的方法为最大标准化残差检验法，该方法检验每个量测量的标准化残差 $r_i^N$，其计算公式如式(3)所示：

$$r_i^N=\frac{\left|z_i-h_i(\hat{\boldsymbol{x}})\right|}{\sqrt{\Omega_{ii}}} \tag{3}$$

其中，$\boldsymbol{\Omega}$ 表示残差协方差矩阵。如果 $r_{\max}^N=\max\{r_i^N \mid i=1, \dots, m\}>c$，则 $r_{\max}^N$ 对应的量测被视为不良数据并将其剔除，然后再用 WLS 进行求解，直至 $r_{\max}^N<c$；$c$ 为给定的阈值，本文中 $c$ 取 3。

图 1 为含 VSC-HVDC 的交直流系统示意图，VSC 通过换流电抗器 $y_{c1}$（$y_{c2}$）和换流变压器 $y_{t1}$（$y_{t2}$）与交流系统相连，本文假设换流器均使用模块化多电平技术，故不考虑交流滤波器。节点 $s_1$、$s_2$ 为 VSC-HVDC 系统与交流系统的连接节点，假定选取交流系统中节点 1 为参考节点，则状态向量 $x^T = [\theta_2\ \theta_3 \ldots \theta_N\ U_1\ U_2 \ldots U_N\ \theta_{c1}\ \theta_{c1}\ U_{c1}\ U_{c2}\ I_{dc1}\ U_{dc1}]$。

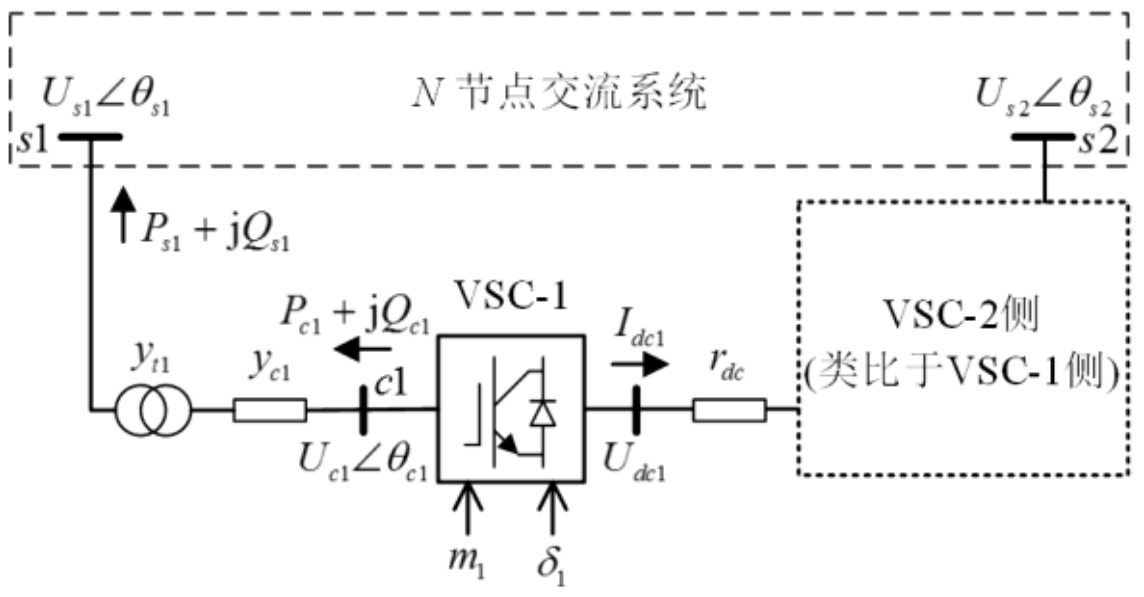

图 1 含 VSC-HVDC 交直流系统示意图

**Fig. 1 Illustration of the AC/DC system with VSC-HVDC**

在交流系统侧，量测量一般包括节点电压幅值、节点注入功率、支路功率等；在 VSC-HVDC 系统侧，量测量一般包括交流侧支路功率、直流侧电压幅值、直流侧电流幅值。交流系统侧和 VSC-HVDC 系统侧量测方程中 $\boldsymbol{h}(\boldsymbol{x})$ 的表达式分别参见文献[19]及附录 A。

### 1.2 VSC-HVDC 系统中换流器的运行限制

为保证换流器的安全运行，VSC-HVDC 系统的稳态运行点必须位于图 2 所示的 P-Q 运行极限图的阴影区域内（下文称之为安全运行区间），以保证流入换流器的交流侧电流幅值和交流侧电压幅值均小于一定的限值[18]。

图 2 中实线圆反映了流入换流器的交流侧电流幅值限制，以换流器 VSC-1 为例，假设允许流入换流器的最大交流侧电流幅值为 $I_{c1,\max}$，则该圆的圆心位于原点，半径为 $U_{s1}I_{c1,\max}$；虚线圆弧反映了换流器交流侧电压幅值限制，假设换流器允许的最大交流侧电压幅值为 $U_{c1,\max}$，虚线圆弧所在圆的圆心为 $(-U_{s1}{}^2 g_{tc1}, U_{s1}{}^2 b_{tc1})$，半径为 $U_{s1}U_{c1,\max}|y_{tc1}|$，其中 $y_{tc1} = g_{tc1} + \mathrm{j}b_{tc1}$ 表示换流电抗器 $y_{c1}$ 和换流变压器 $y_{t1}$ 串联后的等效导纳。

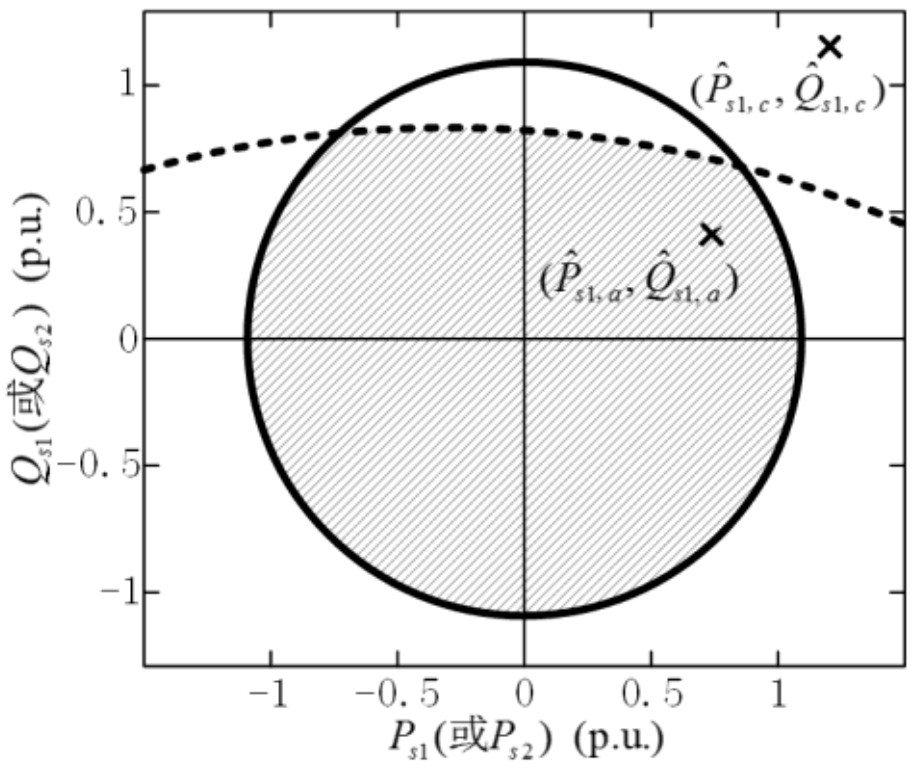

图 2 VSC-HVDC 系统换流器 P-Q 运行极限图

**Fig. 2 P-Q capability chart of the converter station**

## 2　FDI 攻击策略模型的一般形式

如图 3 所示，系统的状态向量、量测向量及相关电气向量分别表示为 $\boldsymbol{x}$、$\boldsymbol{z}$ 及 $\boldsymbol{z}'$，此处的相关电气向量指电网运行人员对电网进行分析、控制及优化时所关心的电气量所构成的向量，如节点电压幅值、支路传输功率、自动发电控制中发电机的目标出力等。假设系统运行于状态 $S_c$，该状态下 $\boldsymbol{x}$、$\boldsymbol{z}$ 及 $\boldsymbol{z}'$ 的真值分别表示为 $\underline{\boldsymbol{x}}_c$、$\underline{\boldsymbol{z}}_c$ 及 $\underline{\boldsymbol{z}}'_c$，电网控制中心获得的量测值向量表示为 $\boldsymbol{z}_c$。SE 根据 $\boldsymbol{z}_c$ 得到系统的状态向量估计值 $\hat{\boldsymbol{x}}_c$，EMS 中的其他高级应用进一步根据 $\hat{\boldsymbol{x}}_c$ 计算得到相关电气量的估计值 $\hat{\boldsymbol{z}}'_c$。需要指出图 3 中的 SE 是指包含了不良数据检测和辨识环节的 SE。

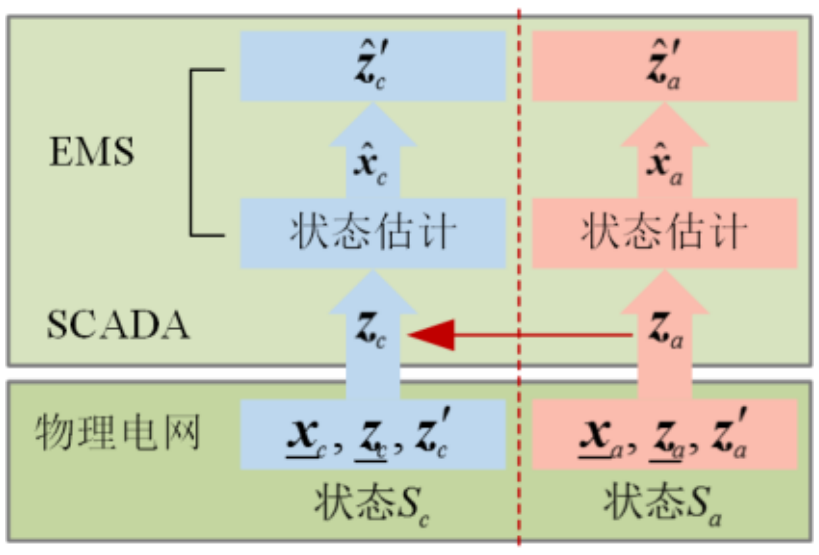

图 3 FDI 攻击示意图

**Fig. 3 Illustration of FDI attacks**

FDI 攻击中攻击者的攻击目的一般有两种：第一种为通过将状态 $S_c$ 下的量测值向量 $\boldsymbol{z}_c$ 篡改为 $\boldsymbol{z}_a$，从而使 SE 的结果变为 $\hat{\boldsymbol{x}}_a$，但并不苛求 $\hat{\boldsymbol{x}}_a$ 的具体取值，只需满足 $\hat{\boldsymbol{x}}_a \neq \hat{\boldsymbol{x}}_c$ 即可，以此为目的的 FDI 攻击称之为随机 FDI 攻击；第二种为通过将状态 $S_c$ 下的量测值向量 $\boldsymbol{z}_c$ 篡改为 $\boldsymbol{z}_a$，从而使得相关电气向量的计算结果变为 $\hat{\boldsymbol{z}}'_a$，且对 $\hat{\boldsymbol{z}}'_a$ 的取值有特定的要求，以此为目的的 FDI 攻击称之为特定 FDI 攻击。此外，攻击者还可能对 $\hat{\boldsymbol{x}}_a$、$\hat{\boldsymbol{z}}'_a$ 甚至 $\boldsymbol{z}_a$ 有更进一步的要求，例如要求 $\|\boldsymbol{z}_a - \boldsymbol{z}_c\|_0$ 小于特定值等，从而降低攻击成本。

简言之，FDI 攻击即为通过将状态 $S_c$ 下的量测值向量 $\boldsymbol{z}_c$ 篡改为 $\boldsymbol{z}_a$，使得 SE 结果和相关电气向量计算结果分别变为 $\hat{\boldsymbol{x}}_a$ 和 $\hat{\boldsymbol{z}}'_a$，且 $\boldsymbol{z}_a$、$\hat{\boldsymbol{x}}_a$ 及 $\hat{\boldsymbol{z}}'_a$ 满足一定的约束条件，从而达到攻击目的。

如何构建 $\boldsymbol{z}_a$ 才能使 SE 根据 $\boldsymbol{z}_a$ 计算得到满足一定约束条件的 $\hat{\boldsymbol{x}}_a$ 呢（对 $\hat{\boldsymbol{z}}'_a$ 的约束条件可转化为对 $\hat{\boldsymbol{x}}_a$ 的约束条件）？如图 3 所示，假设状态向量估计值 $\hat{\boldsymbol{x}}_a$ 对应的系统状态为 $S_a$，该状态下状态向量真值、量测向量真值及相关电气向量真值分别表示为 $\underline{\boldsymbol{x}}_a$、$\underline{\boldsymbol{z}}_a$ 及 $\underline{\boldsymbol{z}}'_a$，若要求 SE 可根据 $\boldsymbol{z}_a$ 计算得到 $\hat{\boldsymbol{x}}_a$，则 $\boldsymbol{z}_a$ 必为系统运行于状态 $S_a$ 时可能出现的量测数据。显然，$\boldsymbol{z}_a$ 可能触发不良数据检测和辨识环节的动作，但只要剔除不良数据后可计算得到 $\hat{\boldsymbol{x}}_a$ 即可。综上，FDI 攻击策略模型的一般形式可表示为以下

优化模型：

$$\boldsymbol{z}_a = \arg\max_{\boldsymbol{z}} f(\hat{\boldsymbol{x}}, \hat{\boldsymbol{z}}', \boldsymbol{z})$$
$$\text{s.t.} \begin{cases} \boldsymbol{z} = \underline{\boldsymbol{z}} + \boldsymbol{e} + \boldsymbol{b} \\ \underline{\boldsymbol{z}} = \boldsymbol{h}(\underline{\boldsymbol{x}}) \\ \hat{\boldsymbol{x}} = f_{SE}(\boldsymbol{z}) \\ \hat{\boldsymbol{z}}' = \boldsymbol{h}'(\hat{\boldsymbol{x}}) \\ \boldsymbol{g}_u(\hat{\boldsymbol{x}}, \hat{\boldsymbol{z}}', \boldsymbol{z}) \le \boldsymbol{0} \\ \boldsymbol{g}_e(\hat{\boldsymbol{x}}, \hat{\boldsymbol{z}}', \boldsymbol{z}) = \boldsymbol{0} \end{cases} \tag{4}$$

其中，$f(\hat{\boldsymbol{x}}, \hat{\boldsymbol{z}}', \boldsymbol{z})$表示与 FDI 攻击中最大化或最小化要求相关的函数；$\underline{\boldsymbol{z}} = \boldsymbol{h}(\underline{\boldsymbol{x}})$反映了$\underline{\boldsymbol{z}}$完全满足系统方程；$f_{SE}(\boldsymbol{z})$为表示 SE 的抽象函数，$\hat{\boldsymbol{x}} = f_{SE}(\boldsymbol{z})$表示 SE 根据量测向量$\boldsymbol{z}$直接得到满意的 SE 结果（含不良数据辨识环节）；$\boldsymbol{h}'(\hat{\boldsymbol{x}})$为 EMS 其他高级应用中计算相关电气向量的函数或抽象函数；$\boldsymbol{g}_u(\hat{\boldsymbol{x}}, \hat{\boldsymbol{z}}', \boldsymbol{z}) \le \boldsymbol{0}$和$\boldsymbol{g}_e(\hat{\boldsymbol{x}}, \hat{\boldsymbol{z}}', \boldsymbol{z}) = \boldsymbol{0}$分别表示 FDI 攻击中的不等式约束和等式约束。当 FDI 攻击中无最大化或最小化要求时，式(4)退化为方程组。

进一步，将式(4)化简可得：

$$\begin{cases} \boldsymbol{z}_a = \boldsymbol{h}(\underline{\boldsymbol{x}}_a) + \boldsymbol{e} + \boldsymbol{b} \\ \underline{\boldsymbol{x}}_a = \arg\max_{\boldsymbol{x}} f(f_{SE}(\boldsymbol{h}(\boldsymbol{x}) + \boldsymbol{e} + \boldsymbol{b}), \\ \qquad \boldsymbol{h}'(f_{SE}(\boldsymbol{h}(\boldsymbol{x}) + \boldsymbol{e} + \boldsymbol{b})), \boldsymbol{h}(\boldsymbol{x}) + \boldsymbol{e} + \boldsymbol{b}) \\ \qquad \text{s.t.} \begin{cases} \boldsymbol{g}_u(f_{SE}(\boldsymbol{h}(\boldsymbol{x}) + \boldsymbol{e} + \boldsymbol{b}), \\ \boldsymbol{h}'(f_{SE}(\boldsymbol{h}(\boldsymbol{x}) + \boldsymbol{e} + \boldsymbol{b})), \boldsymbol{h}(\boldsymbol{x}) + \boldsymbol{e} + \boldsymbol{b}) \le \boldsymbol{0} \\ \boldsymbol{g}_e(f_{SE}(\boldsymbol{h}(\boldsymbol{x}) + \boldsymbol{e} + \boldsymbol{b}), \\ \boldsymbol{h}'(f_{SE}(\boldsymbol{h}(\boldsymbol{x}) + \boldsymbol{e} + \boldsymbol{b})), \boldsymbol{h}(\boldsymbol{x}) + \boldsymbol{e} + \boldsymbol{b}) = \boldsymbol{0} \end{cases} \end{cases} \tag{5}$$

从式(5)不难看出，FDI 攻击策略模型求解的本质是求解状态向量$\underline{\boldsymbol{x}}_a$，即寻找满足一定约束条件的系统状态$S_a$。因此，FDI 攻击策略问题的实质是系统的状态分析问题。

## 3 VSC-HVDC 系统 FDI 攻击策略

攻击者可能对 VSC-HVDC 系统实施随机 FDI 攻击或特定 FDI 攻击，但后者更有针对性和破坏性，因此本文仅考虑针对 VSC-HVDC 系统的特定 FDI 攻击。此外，换流器是整个 VSC-HVDC 系统中最为关键的设备，故本文假设攻击者欲通过实施 FDI 攻击破坏换流器的安全运行。为了达到此攻击目的，攻击策略设计的出发点为通过篡改某些量测量，影响 SE 结果，从而使换流器控制系统对 VSC-HVDC 系统的运行点或安全运行区间做出错误判断。当 VSC-HVDC 系统的真实运行点已超出真实的安全运行区间时，而换流器控制系统中的相应环节并未被触发，这将导致换流器中的换流阀长时间工作于过电流状态或过电压状态而失效，最终使换流器发生损坏[20]。以下以攻击图 1 中换流器 VSC-1 为例对所提出的针对 VSC-HVDC 系统的 FDI 攻击策略进行阐释。

### 3.1 FDI 攻击策略模型

基于第 2 节提出的 FDI 攻击策略模型的一般形式，首先根据攻击目的确定相关电气向量$\boldsymbol{z}' = [P_{s1}\ Q_{s1}]$。如图 2 所示，假设攻击前$\hat{\boldsymbol{z}}_c' = [\hat{P}_{s1,c}\ \hat{Q}_{s1,c}]$，且攻击前换流器 VSC-1 的运行点$(\hat{P}_{s1,c}, \hat{Q}_{s1,c})$位于攻击前安全运行区间以外；攻击后$\hat{\boldsymbol{z}}_a' = [\hat{P}_{s1,a}\ \hat{Q}_{s1,a}]$，且要求攻击后换流器 VSC-1 的运行点$(\hat{P}_{s1,a}, \hat{Q}_{s1,a})$位于攻击后的安全运行区间以内（在电网运行控制人员看来）。另外，考虑到攻击的难度和攻击者掌握的资源限制，攻击者一般对被攻击量测量的数量有一定要求，此处假设攻击者仅要求最小化被攻击量测量的数量（以最小化攻击成本）。从而，根据式(4)可得到针对 VSC-HVDC 系统的 FDI 攻击策略模型如下:

$$\boldsymbol{z}_a = \arg\min_{\boldsymbol{z}} \|\boldsymbol{z} - \boldsymbol{z}_c\|_0$$
$$\text{s.t.} \begin{cases} \boldsymbol{z} = \underline{\boldsymbol{z}} + \boldsymbol{e} + \boldsymbol{b} \\ \underline{\boldsymbol{z}} = \boldsymbol{h}(\underline{\boldsymbol{x}}) \\ \hat{\boldsymbol{x}} = f_{SE}(\boldsymbol{z}) \\ \hat{\boldsymbol{z}}' = \boldsymbol{h}'(\hat{\boldsymbol{x}}) \\ \hat{P}_{s1}{}^2 + \hat{Q}_{s1}{}^2 \le \left(r_1 \hat{U}_{s1} I_{c1,\max}\right)^2 \\ \left(\hat{P}_{s1} + \hat{U}_{s1}{}^2 g_{tc1}\right)^2 + \left(\hat{Q}_{s1} - \hat{U}_{s1}{}^2 b_{tc1}\right)^2 \\ \le \left(r_2 \hat{U}_{s1} U_{c1,\max} |y_{tc1}|\right)^2 \\ \boldsymbol{x}_{\min} \le \hat{\boldsymbol{x}} \le \boldsymbol{x}_{\max} \end{cases} \tag{6}$$

式中，$\|\cdot\|_0$表示向量的 L0 范数；$r_1$、$r_2$为给定的系数，且满足$r_1(\text{或}r_2) \le 1$，当 VSC-HVDC 系统运行中留有一定安全裕度时，其取值需适当减小；$\boldsymbol{x}_{\min}$和$\boldsymbol{x}_{\max}$分别表示系统正常运行状态下状态向量的最小值向量和最大值向量。

### 3.2 FDI 攻击策略模型求解算法

由于式(6)所示的优化模型基于交流 SE 构建、包含 VSC-HVDC 系统特性且同时攻击多个相关电气量，导致已有的 FDI 攻击策略模型求解方法无能为力[8, 13-16]。本小节结合第 2 节对 FDI 攻击策略问题实质的分析，通过将式(6)的求解转化为对状态向量$\underline{x}_a$的求解，亦即寻找系统状态$S_a$，从而给出式(6)的求解算法。另外需要说明的是，当量测数据中含不良数据时，若剔除不良数据后系统仍然可观，则 SE 仍可能得到满意的估计结果。SE 的这一特点可被攻击者用于减小被攻击的量测量的数量，但具体的攻击策略与量测仪表的配置和系统去掉某些量

测量后的可观测性有关。为简化问题，本文不考虑攻击者对这一特性的利用，即令式(6)中 $\boldsymbol{b}=\boldsymbol{0}$。

为便于进一步分析，首先给出以下定义：

● （非）相关状态变量及相关量测量：若量测量 $z_j$（或相关电气量 $z_j'$）是状态变量 $x_i$ 的函数，则称 $x_i$ 为 $z_j$（或 $z_j'$）的相关状态变量，称 $z_j$ 为 $x_i$ 的相关量测量，记作 $x_i \leftrightarrow z_i(z_j')$；否则，称 $x_i$ 为 $z_j$（或 $z_j'$）的非相关状态变量，记作 $x_i \not\leftrightarrow z_i(z_j')$。

● 函数 $\xi$ 及 $\xi'$：

$$\xi(\{z_1,...,z_i\})=\{j\,|\,x_j \leftrightarrow z_1\}\cup...\cup\{j\,|\,x_j \leftrightarrow z_i\} \quad (7)$$

$$\xi'(\{z_1,...,z_i\})=\{j\,|\,x_j \not\leftrightarrow z_1\}\cap...\cap j\,|\,x_j \not\leftrightarrow z_i\} \quad (8)$$

● 向量 $\boldsymbol{w}$： $\boldsymbol{w}=[w_i]_{|z|\times 1}$， $k\in\square^+$， $n_k\in\square^+$。当 $z_i$ 为虚拟量测量时， $w_i=0$，否则 $w_i=1$。

● 集合 $\mathrm{I}_F^{(k,n_k)}$ 及 $\mathrm{I}_V^{(k,n_k)}$：

$$\mathrm{I}_F^k=\begin{cases}\xi'(\{P_{s1},Q_{s1}\}) & k=1,n_k\in\square^+\\ \xi'(\{z_i\,|\,v_i^{(k-1,n_{k-1})}=1\wedge w_i=0\}) & \\ \cap\mathrm{I}_F^1\cap\mathrm{I}_F^2...\cap\mathrm{I}_F^{k-1} & k>1,n_k\in\square^+\end{cases} \quad (9)$$

$$\mathrm{I}_V^k=\begin{cases}\xi(\{P_{s1},Q_{s1}\}) & k=1,n_k\in\square^+\\ \xi(\{z_i\,|\,v_i^{(k-1,n_{k-1})}=1\wedge w_i=0\}) & \\ -(\mathrm{I}_V^1\cup\mathrm{I}_V^2...\cup\mathrm{I}_V^{k-1}) & k>1,n_k\in\square^+\end{cases} \quad (10)$$

● 向量 $\boldsymbol{d}^{(k,n_k)}$： $\boldsymbol{d}^{(k,n_k)}=[d_i^{(k,n_k)}]_{|\mathrm{I}_V^{(k,n_k)}|\times 1}$，其中 $d_i^{(k,n_k)}\in\mathrm{B}$， $k\in\square^+$， $n_k\in\square^+$ 且 $n_k<2^{|\mathrm{I}_V^{(k,n_k)}|}$；对于同一 $k$， $n_k$ 与 $\boldsymbol{d}^{(k,n_k)}$ 所在向量空间 $\mathbf{B}^{|\mathrm{I}_V^{(k,n_k)}|\times 1}$ 中除 $[1]_{|\mathrm{I}_V^{(k,n_k)}|\times 1}$ 外的所有向量一一对应。

● 向量 $\boldsymbol{v}^{(k,n_k)}$： $\boldsymbol{v}^{(k,n_k)}=[v_i^{(k,n_k)}]_{|z|\times 1}$，其中 $k\in\square^+$， $n_k\in\square^+$。若 $k=1$：当 $\exists j\in\mathrm{I}_V^k$ 且 $d_j^{(k,n_k)}=0$，使得 $H_{ij}\neq 0$ 时， $v_i^{(k,n_k)}=1$，否则 $v_i^{(k,n_k)}=0$；若 $k>1$：当 $\exists j\in\mathrm{I}_V^k$ 且 $d_j^{(k,n_k)}=0$，使得 $H_{ij}\neq 0$，且 $i\notin\{i\,|\,v_i^{(k-1,n_{k-1})}=1\wedge w_i=0\}$ 时， $v_i^{(k,n_k)}=1$，否则 $v_i^{(k,n_k)}=0$。其中， $\boldsymbol{H}$ 表示量测方程的雅可比矩阵。

● 集合 $\mathrm{N}_k$： $\mathrm{N}_k=\{n_k^\tau\}$ 为确定的集合，其中 $k\in\square^+$， $n_k^\tau\in\square^+$， $n_k^\tau<2^{|\mathrm{I}_V^k|}$。

对于任意系统状态 $S$， $S=S_a$ 的必要条件为式(11)成立：

$$\begin{cases}\left(-\hat{U}_{s1}{}^2 g_{tc1}+\hat{U}_{s1}\hat{U}_{c1}(g_{tc1}\cos\hat{\theta}_{s1c1}+b_{tc1}\sin\hat{\theta}_{s1c1})\right)^2\\ +\left(\hat{U}_{s1}{}^2 b_{tc1}+\hat{U}_{s1}\hat{U}_{c1}(g_{tc1}\sin\hat{\theta}_{s1c1}-b_{tc1}\cos\hat{\theta}_{s1c1})\right)^2\\ =\left(r_1\hat{U}_{s1}I_{c1,\max}\right)^2\\ \left(-\hat{U}_{s1}{}^2 g_{tc1}+\hat{U}_{s1}\hat{U}_{c1}(g_{tc1}\cos\hat{\theta}_{s1c1}+b_{tc1}\sin\hat{\theta}_{s1c1})+\hat{U}_{s1}{}^2 g_{tc1}\right)^2\\ +\left(\hat{U}_{s1}{}^2 b_{tc1}+\hat{U}_{s1}\hat{U}_{c1}(g_{tc1}\sin\hat{\theta}_{s1c1}-b_{tc1}\cos\hat{\theta}_{s1c1})-\hat{U}_{s1}{}^2 b_{tc1}\right)^2\\ =\left(r_2\hat{U}_{s1}U_{c1,\max}|y_{tc1}|\right)^2\\ \boldsymbol{x}_{\min}\le\hat{\boldsymbol{x}}\le\boldsymbol{x}_{\max}\end{cases} \quad (11)$$

由于攻击前的运行点 $(\hat{P}_{s1,c},\hat{Q}_{s1,c})$ 位于攻击前的安全运行区间以外，故当 $S=S_c$ 时，式(11)必不成立；则当假设虚拟量测量可被攻击时，状态 $S_a$ 与 $S_c$ 的差异必然存在于且仅存在于 $P_{s1}$ 和 $Q_{s1}$ 的相关状态变量。进而， $S=S_a$ 的必要条件除式(11)外，还需有下式成立：

$$O_1(n_1^1)\vee O_1(n_1^2)...\vee O_1(n_1^\tau)...\vee O_1(n_1^{|\mathrm{N}_1|}) \quad n_1^\tau\in\mathrm{N}_1 \quad (12)$$

其中， $O_1(n_1^\tau)$ 表示以下方程组：

$$\begin{cases}\underline{x}_i-\underline{x}_{c,i}=0 & i\in\mathrm{I}_F^1\\ d_i^{(1,n_1^\tau)}\left(\underline{x}_i-\underline{x}_{c,i}\right)=0 & i\in\mathrm{I}_V^1\end{cases} \quad (13)$$

式(12)与式(11)联立即构成了 $S=S_a$ 成立的充分必要条件，考虑到联立式(12)与式(11)可能得到多种系统状态，即 $S_a$ 并不唯一，为简化问题，本文以状态变量的偏移量最小为原则确定唯一解。另外，在式(11)中，可近似认为 $\hat{\boldsymbol{x}}=\underline{\boldsymbol{x}}$，在式(12)中，可近似认为 $\underline{\boldsymbol{x}}_c=\hat{\boldsymbol{x}}_c$。则式(12)与式(11)联立后可转化为以下优化问题：

$$\underline{\boldsymbol{x}}_a=\arg\min_{\underline{\boldsymbol{x}},n_1^\tau}\|\underline{\boldsymbol{x}}-\hat{\boldsymbol{x}}_c\|_2$$

$$\text{s.t.}\begin{cases}\underline{x}_i-\hat{x}_{c,i}=0 & i\in\mathrm{I}_F^1\\ d_i^{(1,n_1^\tau)}\left(\underline{x}_i-\hat{x}_{c,i}\right)=0 & i\in\mathrm{I}_V^1\\ \boldsymbol{g}_{z'}(\underline{\boldsymbol{x}})=\boldsymbol{0}\\ n_1^\tau\in\mathrm{N}_1\end{cases} \quad (14)$$

式中， $\|\bullet\|_2$ 表示向量的 L2 范数， $\boldsymbol{g}_{z'}(\underline{\boldsymbol{x}})=\boldsymbol{0}$ 表示式(11)中状态变量的估计值用状态变量的真值替换后得到的方程组。

通过求解式(14)得到 $S_a$ 的前提条件是虚拟量测量可被攻击，而事实并非如此。当式(14)中 $n_1^\tau$ 取 $\mathrm{N}_1$ 中的某一元素 $n_1$ 时，则

1) 当 $\boldsymbol{v}^{(1,n_1)T}\boldsymbol{w}=\boldsymbol{v}^{(1,n_1)T}\boldsymbol{v}^{(1,n_1)}$ 时， $\forall i\in\mathrm{I}_V^1$ 且 $d_i^{(1,n_1)}=0$， $x_i$ 的相关量测量均非虚拟量测量，此时 $\|\boldsymbol{z}-\boldsymbol{z}_c\|_0=\|\boldsymbol{v}^{(1,n_1)}\times\boldsymbol{w}\|_0$，攻击者仅需攻击 $x_i$ 的所有相关量测量，即 $\{z_j\,|\,z_j\leftrightarrow x_i, i\in\mathrm{I}_V^{(1,n_1)}\wedge d_i^{(1,n_1)}=0\}$。

2) 当 $\boldsymbol{v}^{(1,n_1)T}\boldsymbol{w}<\boldsymbol{v}^{(1,n_1)T}\boldsymbol{v}^{(1,n_1)}$ 时， $\exists i\in\mathrm{I}_V^1$ 且 $d_i^{(1,n_1)}=0$，使得 $x_i$ 的相关量测量中含有虚拟量测量，由于虚拟量测量本质为系统约束而无法进行攻击，此时式(14)变为:

$$\underline{\boldsymbol{x}}_a=\arg\min_{\underline{\boldsymbol{x}}}\|\underline{\boldsymbol{x}}-\hat{\boldsymbol{x}}_c\|_2$$

$$\text{s.t.}\begin{cases}\underline{x}_i-\underline{\hat{x}}_{c,i}=0 & i\in\mathrm{I}_F^1\\ d_i^{(1,n_1)}\left(\underline{x}_i-\hat{x}_{c,i}\right)=0 & i\in\mathrm{I}_V^1\\ \boldsymbol{g}_{z'}(\underline{\boldsymbol{x}})=\boldsymbol{0}\\ z_i=h_i(\underline{x}) & i\in\{i\,|\,v_i^{(1,n_1)}=1\wedge w_i=0\}\end{cases} \quad (15)$$

若式(15)存在可行解，则攻击者仅需攻击 $x_i$ 的所有相关量测量；若式(15)无可行解，则当假设除 $x_i$（$i\in \mathrm{I}_V^1$ 且 $d_i^{(1,n_1)}=0$）的相关量测量中含有的虚拟量测量外，其余虚拟量测量均可被攻击时，式(14)变为：

$$\underline{\boldsymbol{x}}_a=\underset{\underline{\boldsymbol{x}},\,n_2^{\tau}}{\arg\min}\ \|\underline{\boldsymbol{x}}-\hat{\underline{\boldsymbol{x}}}_c\|_2$$
$$\text{s.t.}\begin{cases}\underline{x}_i-\hat{\underline{x}}_{c,i}=0 & i\in \mathrm{I}_F^2\\ d_i^{(1,n_1)}\left(\underline{x}_i-\hat{\underline{x}}_{c,i}\right)=0 & i\in \mathrm{I}_V^1\\ d_i^{(2,n_2^{\tau})}\left(\underline{x}_i-\hat{\underline{x}}_{c,i}\right)=0 & i\in \mathrm{I}_V^2\\ \boldsymbol{g}_{z'}(\underline{\boldsymbol{x}})=\boldsymbol{0}\\ z_i=h_i(\underline{x}) & i\in\{i\mid v_i^{(1,n_1)}=1\wedge w_i=0\}\\ n_2^{\tau}\in \mathrm{N}_2\end{cases}\tag{16}$$

同理式(14)可对式(16)进行进一步分析计算。以上求解过程实施的困难在于集合 $\mathrm{N}_k$ 中的元素未知，一个可行的做法是假设 $\mathrm{N}_k$ 中仅含有一个元素，使该元素遍历其所有可能的取值，求解后得到令 $\|\boldsymbol{z}-\boldsymbol{z}_c\|_0$ 最小的 $\mathrm{N}_k$ 中的元素即构成真正的集合 $\mathrm{N}_k$。遍历过程中已经得到 $\mathrm{N}_k$ 中各元素对应的使 $\|\underline{\boldsymbol{x}}-\hat{\underline{\boldsymbol{x}}}_c\|_2$ 最小化的系统状态，假设为 $\{\underline{\boldsymbol{x}}_a^t\mid t=1,2,\ldots\}$，则最终的系统状态 $\underline{\boldsymbol{x}}_a$ 确定为：

$$\underline{\boldsymbol{x}}_a=\underset{\underline{\boldsymbol{x}}_a^t}{\arg\min}\ \|\underline{\boldsymbol{x}}_a^t-\hat{\underline{\boldsymbol{x}}}_c\|_2\tag{17}$$

显然穷举遍历可能无法满足实施FDI攻击对计算时间的要求，攻击者完全有可能利用一定的遍历规则减小求解时间，但本文的目的在于对VSC-HVDC 系统 FDI 攻击的存在性及其特点进行分析，并不深入讨论求解时间的问题。为使计算时间在本文的研究目的下可被接受，采用以下方法初步减小遍历时间：

以遍历 $\mathrm{N}_1$ 中的元素至 $n_i$ 为例，假设之前遍历过程中所得到的需要篡改的量测量数目的最小值为 $\overline{\ell}$，显然 $\mathrm{N}_1$ 取元素 $n_i$ 时需要篡改的量测量数目必然大于或等于 $\|\boldsymbol{v}^{(1,n_i)}\times\boldsymbol{w}\|_0$，故若 $\|\boldsymbol{v}^{(1,n_i)}\times\boldsymbol{w}\|_0>\overline{\ell}$，则不对 $\mathrm{N}_1$ 取元素 $n_i$ 的情况做进一步分析。

基于上述分析，给出模型的求解算法步骤如图4 所示，图 4 中 $M$ 为一个充分大的正整数，$\ell_t=\|\boldsymbol{w}\times\sum_{j=1}^{k}\boldsymbol{v}^{(j,n_j)}\|_0$ 表示需要篡改的量测量的数量，$Z_t$ 表示需要攻击的量测量集合，且 $Z_t=\{z_j\mid z_j\leftrightarrow x_i, i\in \mathrm{I}_V^{(l,n_l)}\wedge d_i^{(l,n_l)}=0\wedge w_j=1, l\in \square^{+}\wedge l\le k\}$。图 4 中式(18)-式(20)的表达式如下：

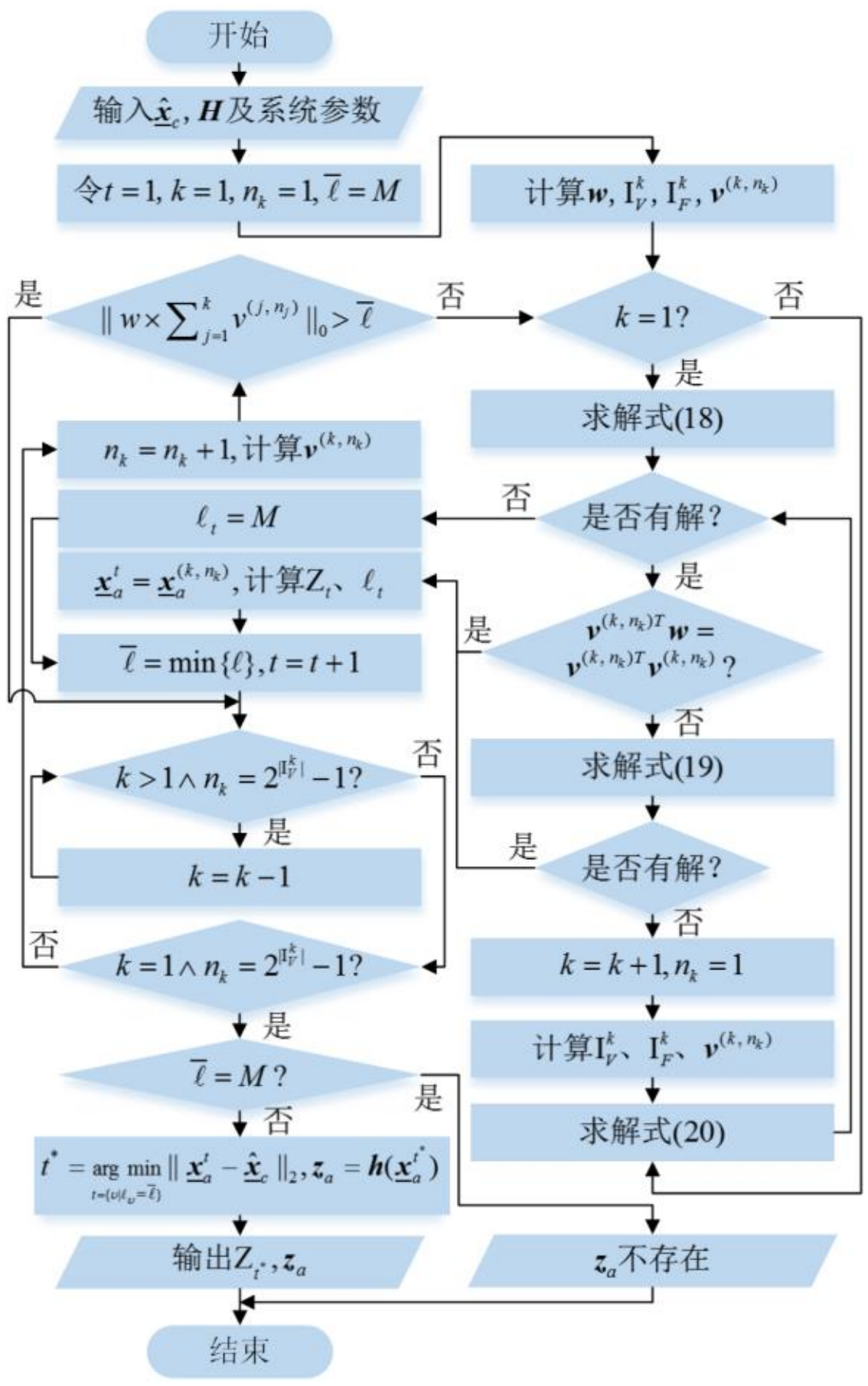

图 4　FDI 攻击策略模型求解算法流程图

**Fig. 4　Flow chart of the algorithm for solving the model of FDI attack strategies**

$$\underline{\boldsymbol{x}}_a=\underset{\underline{\boldsymbol{x}}}{\arg\min}\ \|\underline{\boldsymbol{x}}-\hat{\underline{\boldsymbol{x}}}_c\|_2$$
$$\text{s.t.}\begin{cases}\underline{x}_i-\hat{\underline{x}}_{c,i}=0 & i\in \mathrm{I}_F^1\\ d_i^{(1,n_1)}\left(\underline{x}_i-\hat{\underline{x}}_{c,i}\right)=0 & i\in \mathrm{I}_V^1\\ \boldsymbol{g}_{z'}(\underline{\boldsymbol{x}})=\boldsymbol{0}\end{cases}\tag{18}$$

$$\underline{\boldsymbol{x}}_a=\underset{\underline{\boldsymbol{x}}}{\arg\min}\ \|\underline{\boldsymbol{x}}-\hat{\underline{\boldsymbol{x}}}_c\|_2$$
$$\text{s.t.}\begin{cases}\underline{x}_i-\hat{\underline{x}}_{c,i}=0 & i\in \mathrm{I}_F^k\\ d_i^{(l,n_l)}\left(\underline{x}_i-\hat{\underline{x}}_{c,i}\right)=0 & i\in \mathrm{I}_V^l, l\in \square^{+}\wedge l\le k\\ \boldsymbol{g}_{z'}(\underline{\boldsymbol{x}})=\boldsymbol{0}\\ z_i=h_i(\underline{x}) & i\in\{i\mid (v_i^{(1,n_1)}=1\vee v_i^{(2,n_2)}=1\\ & \quad\vee\ldots\vee v_i^{(k,n_k)}=1)\wedge w_i=0\}\end{cases}\tag{19}$$

$$\underline{\boldsymbol{x}}_a=\underset{\underline{\boldsymbol{x}}}{\arg\min}\ \|\underline{\boldsymbol{x}}-\hat{\underline{\boldsymbol{x}}}_c\|_2$$
$$\text{s.t.}\begin{cases}\underline{x}_i-\hat{\underline{x}}_{c,i}=0 & i\in \mathrm{I}_F^k\\ d_i^{(l,n_l)}\left(\underline{x}_i-\hat{\underline{x}}_{c,i}\right)=0 & i\in \mathrm{I}_V^l, l\in \square^{+}\wedge l\le k\\ \boldsymbol{g}_{z'}(\underline{\boldsymbol{x}})=\boldsymbol{0}\\ z_i=h_i(\underline{x}) & i\in\{i\mid (v_i^{(1,n_1)}=1\vee v_i^{(2,n_2)}=1\\ & \quad\vee\ldots\vee v_i^{(k-1,n_{k-1})}=1)\wedge w_i=0\}\end{cases}\tag{20}$$

## 4　算例分析

本节利用修改的 IEEE-14 节点系统对所提出的针对 VSC-HVDC 系统的 FDI 攻击策略的有效性进行验证。

### 4.1 算例参数

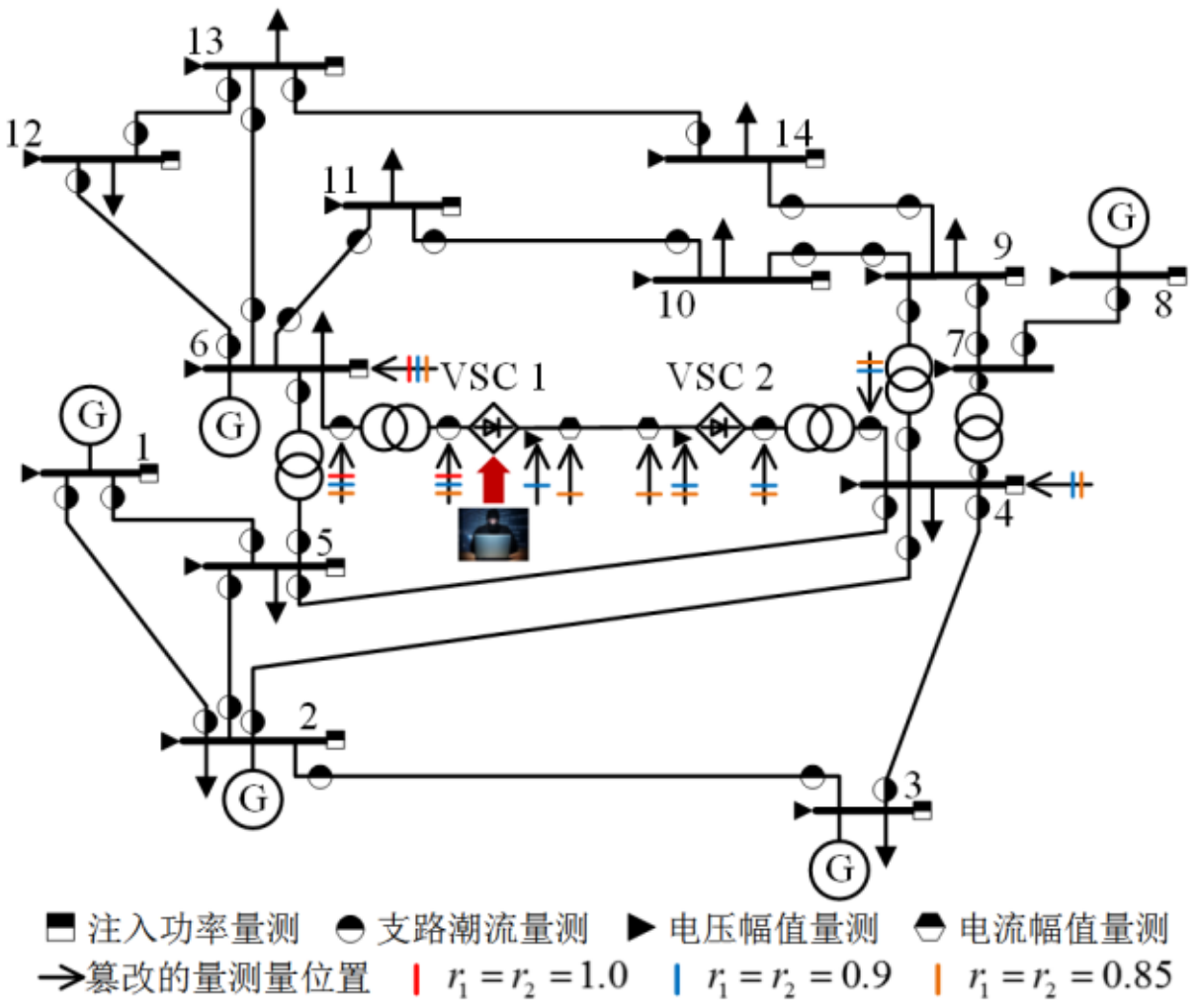

图 5　修改的 IEEE-14 节点系统单线图

**Fig. 5　The diagram of the modified IEEE-14 bus system**

如图 5 所示，在原 IEEE-14 节点系统的节点 6 和节点 4 之间增加一条 VSC-HVDC 支路，其参数参见附录 B；系统量测配置为：所有交流节点配备电压幅值量测，所有非零功率注入节点配备注入功率量测，所有交流支路（包含换流变压器）配备首末端潮流量测，各换流器配备直流侧电压幅值量测、直流侧电流幅值量测。量测值通过在 FDI 攻击前量测真值的基础上叠加一定的高斯噪声获得，各量测仪表均取 $\sigma=10^{-3}$。FDI 攻击前量测真值根据如附录 C 所示的 FDI 攻击前系统的运行状态计算得到。

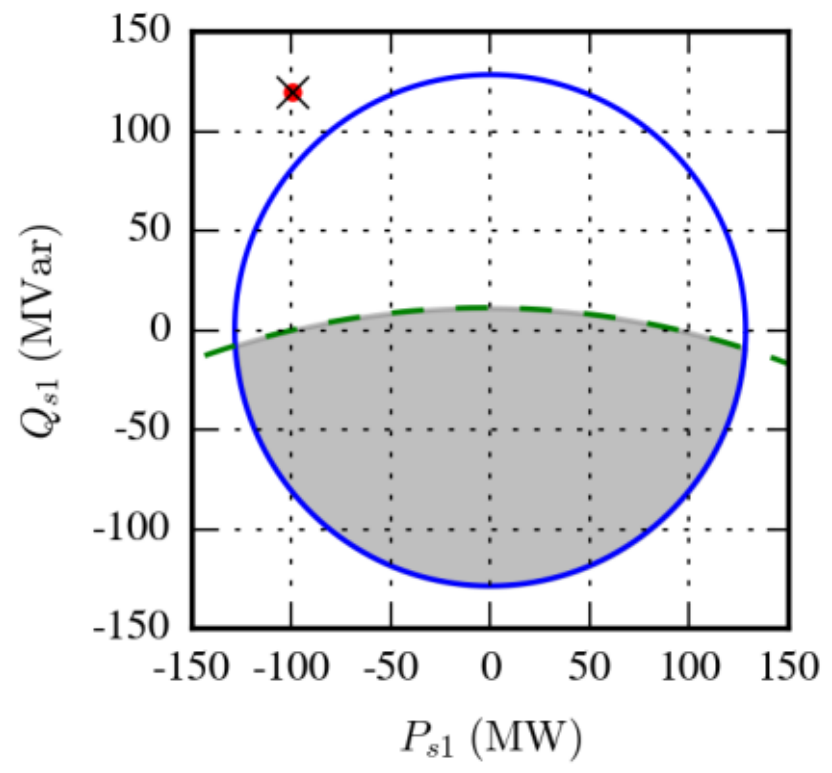

图 6　FDI 攻击前换流器 VSC-1 的 P-Q 运行极限及运行点

**Fig. 6　P-Q capability chart and the operation point of the converter station VSC-1 before FDI attacks**

另外，FDI 攻击前换流器 VSC-1 的运行点如图 6 中红色圆点所示。此时，换流器 VSC-1 的运行点已经大幅度超出其安全运行区间（图 6 中的灰色阴影区域）。正常情况下，换流器的控制系统将会动作使运行点位于安全运行区间以内。

### 4.2 算例结果

为验证所提攻击策略的有效性，在当前的系统运行状态下，分别取 $r_1=r_2=1.0$、$r_1=r_2=0.9$、$r_1=r_2=0.85$ 三组不同的取值，每组取值下进行 100 次独立试验，且每次试验均满足 FDI 攻击前 $r_{\max}^N \le c$。另外，不同组之间的 100 次独立试验产生的量测值彼此相同。

表 1　FDI 攻击成功率及需要篡改的量测量数目

**Tab.1　Success rate of FDI attacks and the number of tampered measurements**

| 组别 | $r_1=r_2=1.0$ | $r_1=r_2=0.9$ | $r_1=r_2=0.85$ |
|---|---|---|---|
| 成功率 | 94% | 94% | 97% |
| 需要篡改的量测量数目 | 6 | 14 | 15 |

表 1 给出了三组不同的 $r_1$、$r_2$ 取值下实施 FDI 攻击的成功率及需要篡改的量测数目，具体需要篡改的量测量如图 5 所示，攻击前后的量测值如图 7 所示（为清晰起见，在不同的 $r_1$、$r_2$ 取值下仅分别给出某一次试验下的结果，且攻击前的量测值彼此相同；若某 $r_1$、$r_2$ 取值下量测量的量测值未标出，则表示该量测量在该 $r_1$、$r_2$ 取值下未被篡改）。需要说明，若 FDI 攻击后 $r_{\max}^N<3$，则认为本次 FDI 攻击成功，否则认为本次 FDI 攻击失败；另外，每组取值下不同的独立试验中需要篡改的量测量数目相同。从表中可以看出，三组不同的 $r_1$、$r_2$ 取值下，FDI 攻击的成功率均大于或等于 94%，且最少仅需篡改 6 个量测量即可成功地对换流器 VSC-1 实施 FDI 攻击。

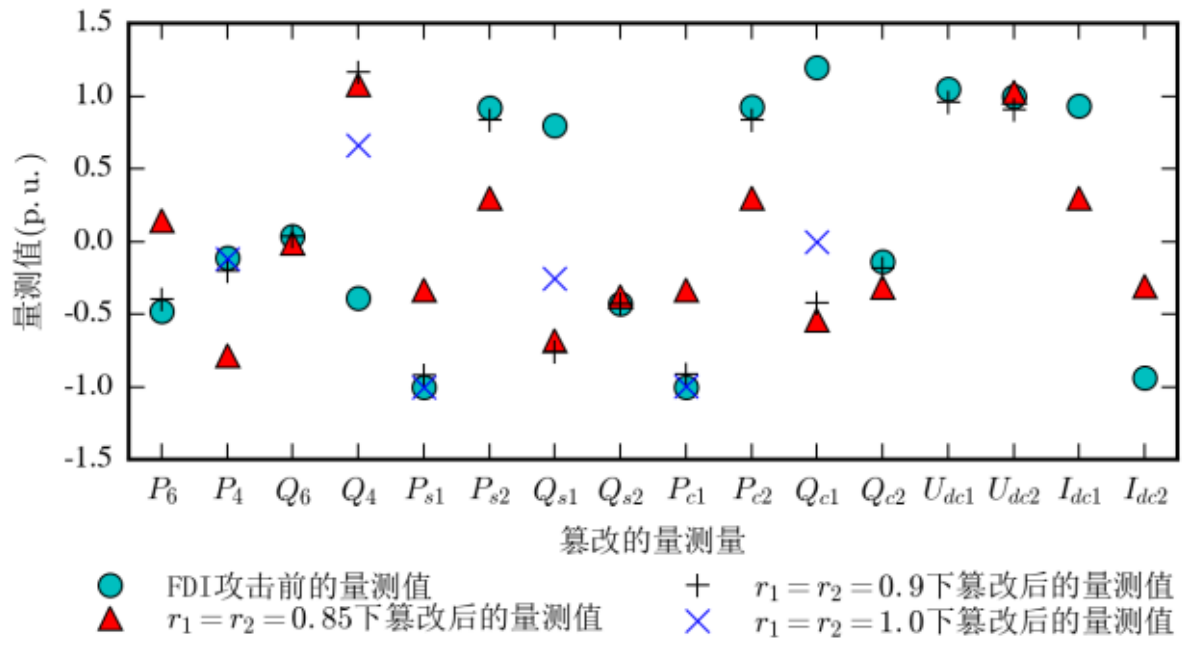

图 7　FDI 攻击前后的量测值（仅给出被篡改的量测量）

**Fig. 7　Measurement values before and after FDI attacks (only those being tampered are listed)**

图 8 给出了三组不同的 $r_1$、$r_2$ 取值下实施 FDI 攻击前后最大标准化残差 $r_{\max}^N$ 的对比情况。从图中可以看出，FDI 攻击均未引起 $r_{\max}^N$ 的明显增大，且在大部分试验下，FDI 攻击后的 $r_{\max}^N$ 较攻击前有不同程度的减小。一方面，由于 FDI 攻击后量测向量 $\boldsymbol{z}_a$ 中经篡改的量测量是将状态向量 $\underline{\boldsymbol{x}}_a$ 代入 $\boldsymbol{h}(\underline{\boldsymbol{x}})$ 中计算而得，使得量测向量 $\boldsymbol{z}_a$ 的一致性较 FDI 攻击前的量测值向量 $\boldsymbol{z}_c$ 更好，从而导致 FDI 攻击后的 $r_{\max}^N$ 较攻击前有所减小；而另一方面，求解 $\boldsymbol{z}_a$ 过程中用 $\hat{\underline{\boldsymbol{x}}}_c$ 近似替代 $\underline{\boldsymbol{x}}_c$ 会降低量测向量 $\boldsymbol{z}_a$ 的一致性，从而可能出现 FDI 攻击后的 $r_{\max}^N$ 较攻击前略微增大的情况。

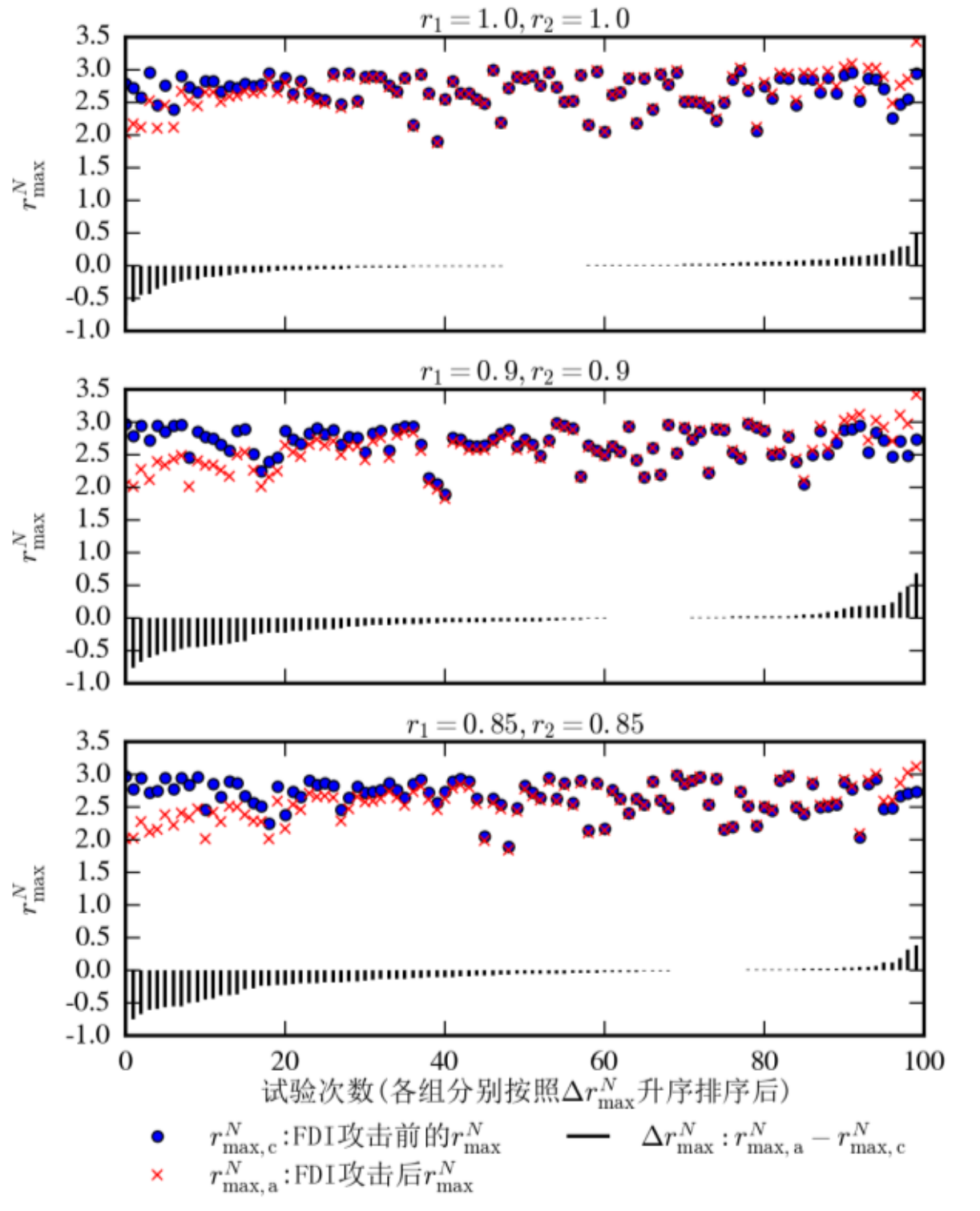

图 8　FDI 攻击后的最大标准化残差

**Fig. 8　Values of the largest normalized residual under FDI attacks**

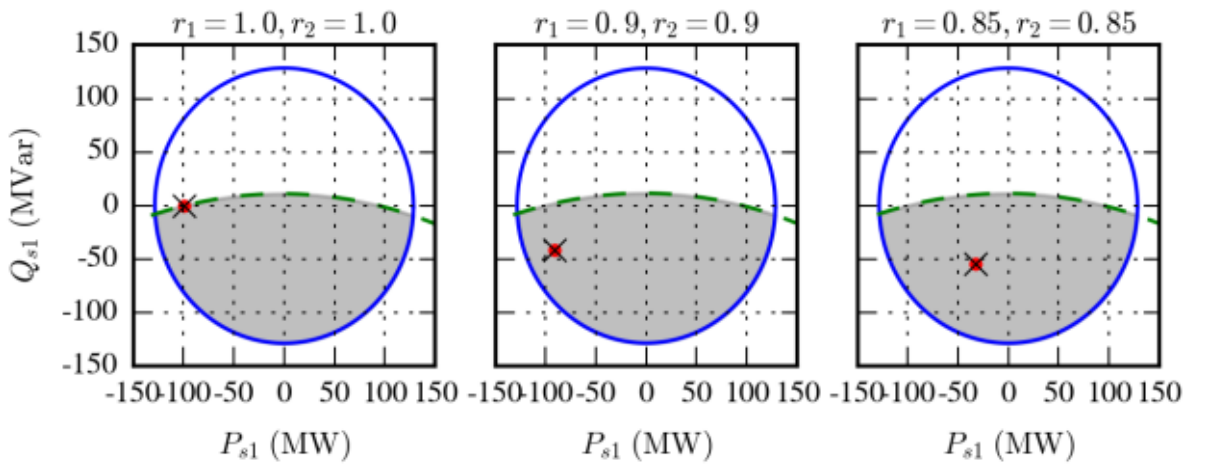

图 9　FDI 攻击后换流器 VSC-1 的 P-Q 运行极限及运行点

**Fig. 9　P-Q capability chart and the operation point of the converter station VSC-1 under FDI attacks**

图 9 为 FDI 攻击后根据 SE 结果计算得到的换流器 VSC-1 的安全运行区间（图中灰色阴影区域）及运行点（图中红色圆点）。在同一 $r_1$、$r_2$ 取值下，图 9 中仅给出某一次试验下的结果，实际上对于同一 $r_1$、$r_2$ 取值下的不同试验，P-Q 运行极限及运行点位置的差别极小。结合图 6 和图 9 可知，换流器 VSC-1 的真实运行点已经大幅度超出其安全运行区间，但 FDI 攻击之后，根据 SE 结果计算得到的换流器 VSC-1 的运行点均位于安全运行区间内，从而使换流器控制系统对换流器的运行状态做出错误判断。这将导致换流器中换流阀长时间遭受过电压和过电流而失效，最终使换流器发生损坏，而 VSC-HVDC 系统故障后发生的直流功率转移甚至可能进一步威胁交流系统的安全稳定运行。

### 4.3 量测配置对 FDI 攻击的影响

在实际电网中，量测配置一般无法满足 4.1 小节中设定的满量测配置，因此，有必要进一步分析量测配置对 FDI 攻击的影响。本小节在 8 组不同的量测配置下，分别重复 4.2 小节中的试验，除量测配置外，其余参数均与 4.1 小节中相同。8 组量测配置中，第 1 组即为 4.1 小节中设定的量测配置，第 2 组在第 1 组的基础上通过去掉交流系统中的某些量测使量测冗余度降低，第 3~8 组分别在前一组的基础上通过去掉 VSC-HVDC 系统（包含与交流系统的连接节点）中的某些量测使量测冗余度依次降低，详细的量测配置参见附录 D。

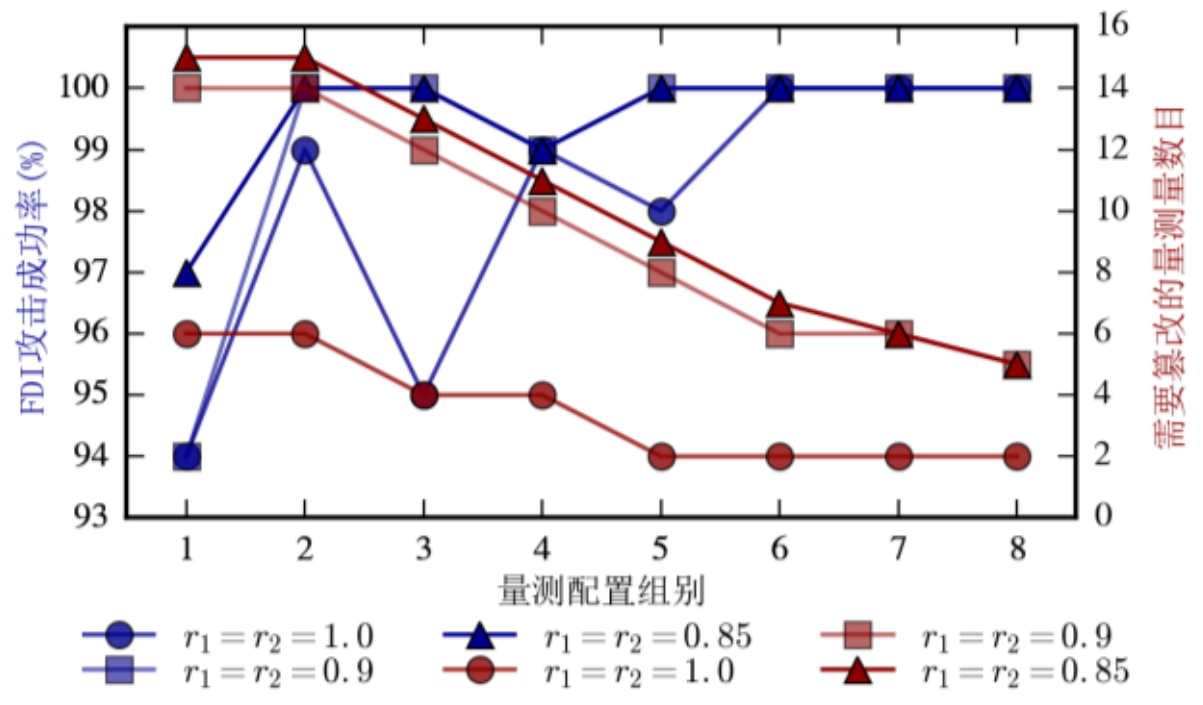

图 10　不同量测配置下 FDI 攻击成功率及需要篡改的量测量数目

**Fig. 10　Success rates of FDI attacks and numbers of measurements being tampered**

图 10 给出了在第 1~8 组量测配置下 FDI 攻击成功率及需要篡改的量测量数目。从图中可以看出，在不同的 $r_1$、$r_2$ 取值下，第 2~8 组的 FDI 攻击成功率均高于第 1 组，而需要篡改的量测量数目均小于第 1 组。在第 8 组量测配置下，仅需要篡改 2 个或 5 个量测量即可成功对 VSC-HVDC 实施 FDI 攻击。因此，当量测配置为非满量测配置时，攻击者完全可能通过修改更少的量测量，而以更高的成功率对 VSC-HVDC 系统实施 FDI 攻击。

## 5　结论

本文提出了电力系统 FDI 攻击策略模型的一般形式，并对 FDI 攻击策略问题的实质进行了分析。进一步提出了以破坏 VSC-HVDC 系统换流器运行安全为目标的 FDI 攻击策略模型及求解方法。最后以修改的 IEEE14 节点系统为例展示了 VSC-HVDC

系统所面临的 FDI 攻击问题，结果表明攻击者通过篡改较少数量的量测仪表即可 VSC-HVDC 系统成功实施 FDI 攻击，从而使 VSC-HVDC 系统因遭受 FDI 攻击而长时间运行于安全运行区间以外，从而导致换流器因遭受过电压及过电流而发生损坏，并可能进一步威胁交流系统的安全稳定运行。另外，本文所提出的电力系统FDI攻击策略模型的一般形式和针对 VSC-HVDC 系统的 FDI 攻击策略模型的求解方法对于电力系统其他形式的FDI攻击问题的研究亦具有指导和借鉴意义。

## 参考文献

### 附录 A：VSC-HVDC 系统侧的量测方程

VSC-HVDC 系统侧量测方程中的 $h(x)$ 由式(A-1)-式(A-13)给出（对于仅给出换流器 VSC-1 侧的 $h(x)$ 可类比至换流器 VSC-2 侧）：

$$P_{s1} = -U_{s1}{}^2 g_{tc1} + U_{s1}U_{c1}(g_{tc1}\cos\theta_{s1c1} + b_{tc1}\sin\theta_{s1c1}) \quad \text{(A-1)}$$

$$Q_{s1} = U_{s1}{}^2 b_{tc1} + U_{s1}U_{c1}(g_{tc1}\sin\theta_{s1c1} - b_{tc1}\cos\theta_{s1c1}) \quad \text{(A-2)}$$

$$P_{c1} = U_{c1}{}^2 g_{tc1} - U_{c1}U_{s1}(g_{tc1}\cos\theta_{c1s1} + b_{tc1}\sin\theta_{c1s1}) \quad \text{(A-3)}$$

$$Q_{c1} = -U_{c1}{}^2 b_{tc1} - U_{c1}U_{s1}(g_{tc1}\sin\theta_{c1s1} - b_{tc1}\cos\theta_{c1s1}) \quad \text{(A-4)}$$

$$U_{dc1} = U_{dc1} \quad \text{(A-5)}$$

$$I_{dc1} = I_{dc1} \quad \text{(A-6)}$$

$$U_{dc2} = U_{dc1} - I_{dc1}r_{dc} \quad \text{(A-7)}$$

$$I_{dc2} = -I_{dc1} \quad \text{(A-8)}$$

此外，换流器还需满足如下有功功率平衡约束：

$$P_{loss1} + P_{c1} + P_{dc1} = 0 \quad \text{(A-9)}$$

其中，$P_{dc1}$ 表示换流器 VSC-1 的直流功率，其计算公式如下：

$$P_{dc1} = U_{dc1}I_{dc1} \quad \text{(A-10)}$$

$P_{loss1}$ 表示换流器VSC-1的有功损耗，其计算公式如下：

$$P_{loss1} = a_1 + b_1 I_{dc1} + c_1 I_{dc1}^{\ 2} \tag{A-11}$$

其中，$a_1$、$b_1$、$c_1$ 为 换流器 VSC-1 的损耗系数，对换流器工作于整流状态和逆变状态两种情况，$c_1$ 的取值不同；$I_{dc1}$ 表示换流器 VSC-1 交流测的电流幅值，其计算公式如下：

$$I_{dc1} = \sqrt{(g_{tc1} + b_{tc1})(U_{c1}^{\ 2} + U_{s1}^{\ 2} - 2U_{c1}U_{s1}\cos\theta_{c1s1})} \tag{A-12}$$

在换流器 VSC-2 的有功功率平衡约束中，$P_{dc2}$ 应按下式计算：

$$P_{dc2} = -(U_{dc1} - I_{dc1} r_{dc}) I_{dc1} \tag{A-13}$$

## 附录 B：VSC-HVDC 系统参数

表 B1　VSC-HVDC 系统参数

**Tab.B1　Parameters of the VSC-HVDC system**

| 参数 | 取值 |
| --- | --- |
| $y_{t1}$ /(°) | 0.119-8.919j |
| $y_{c1}$ /(°) | 0.0037-6.087j |
| $a_1$ | 0.0037-6.087j |
| $b_1$ | 0.887 |
| $c_1$ (整流) | 2.885 |
| $c_1$ (逆变) | 4.371 |
| $I_{c1,max}$ / (p. u. ) | 1.2 |
| $U_{c1,max}$ /(p. u. ) | 1.1 |
| $r_{dc}$ | 0.052 |

注：交流侧与直流侧电压基准值均为 345KV，功率基准值均为100MVA(下同)；换流器 VSC-2 与换流器 VSC-1 参数相同。

## 附录 C：FDI 攻击前系统的运行状态

表 C1　VSC-HVDC 侧状态变量取值

**Tab.C1　Values of state variables in the VSC-HVDC system**

| 参数 | 取值 |
| --- | --- |
| $\theta_{c1}$ /(°) | -34.993 |
| $\theta_{c2}$ /(°) | 6.397 |
| $U_{c1}$ / (p. u. ) | 1.301 |
| $U_{c2}$ / (p. u. ) | 0.920 |
| $I_{dc1}$ / (p. u. ) | 0.937 |
| $U_{dc1}$ / (p. u. ) | 1.049 |

表 C2　交流系统侧状态变量取值

**Tab.C2　Values of state variables in the AC system**

| 交流节点编号 | 电压幅值/(p. u. ) | 电压角度/(°) |
| --- | --- | --- |
| 1 | 1.060 | 0.000* |
| 2 | 1.045 | -5.089 |
| 3 | 1.010 | -12.707 |
| 4 | 1.000 | -9.727 |
| 5 | 1.000 | -9.479 |
| 6 | 1.070 | -23.490 |
| 7 | 1.057 | -15.319 |
| 8 | 1.090 | -15.319 |
| 9 | 1.058 | -18.158 |
| 10 | 1.053 | -19.375 |
| 11 | 1.058 | -21.518 |
| 12 | 1.054 | -23.927 |
| 13 | 1.051 | -23.568 |
| 14 | 1.037 | -21.502 |

注：电压基准值为 345KV，功率基准值为 100MVA。

## 附录 D：第 1~8 组量测配置

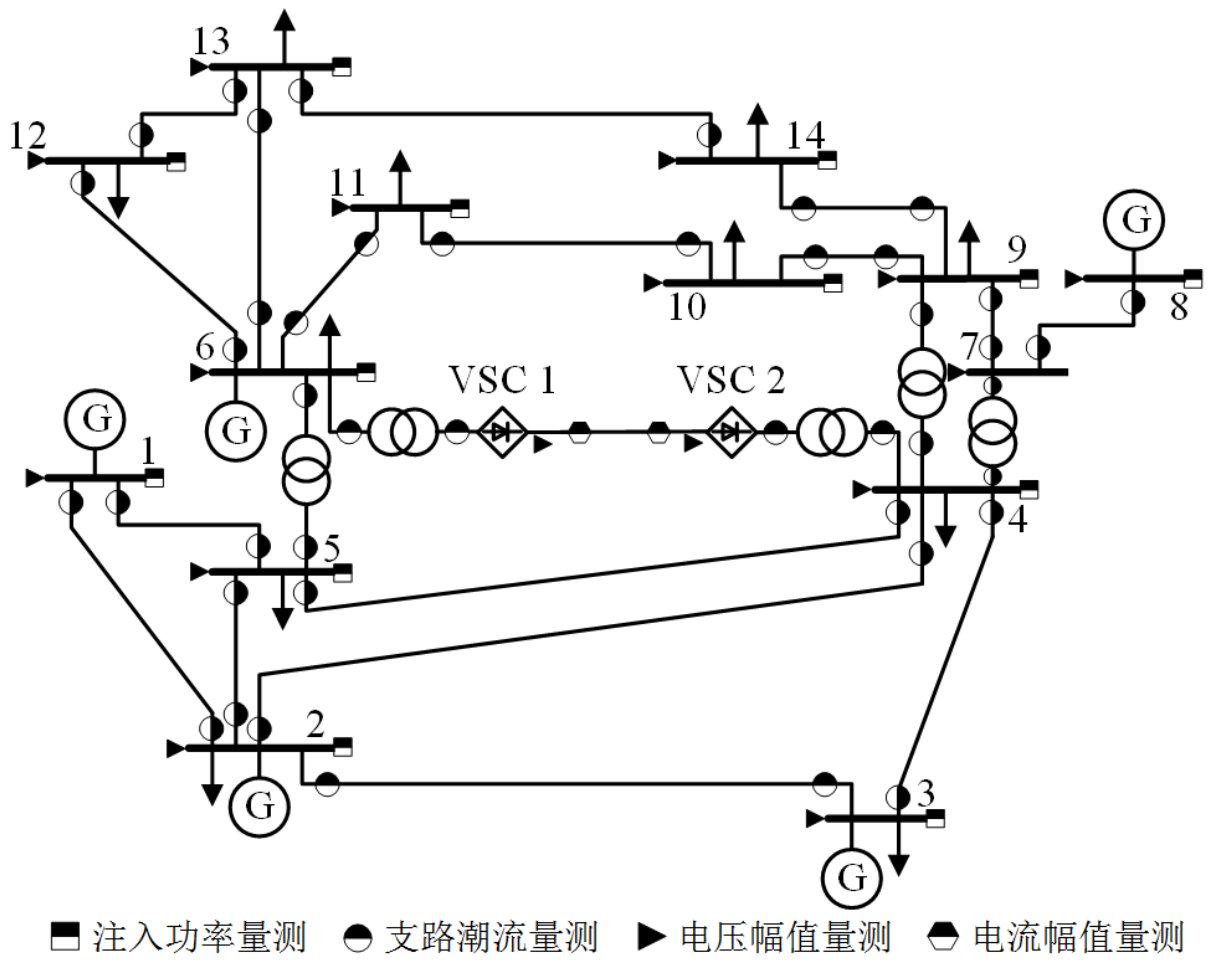

图 D1　第 1 组量测配置

**Fig. D1　Measurement configurations of group 1**

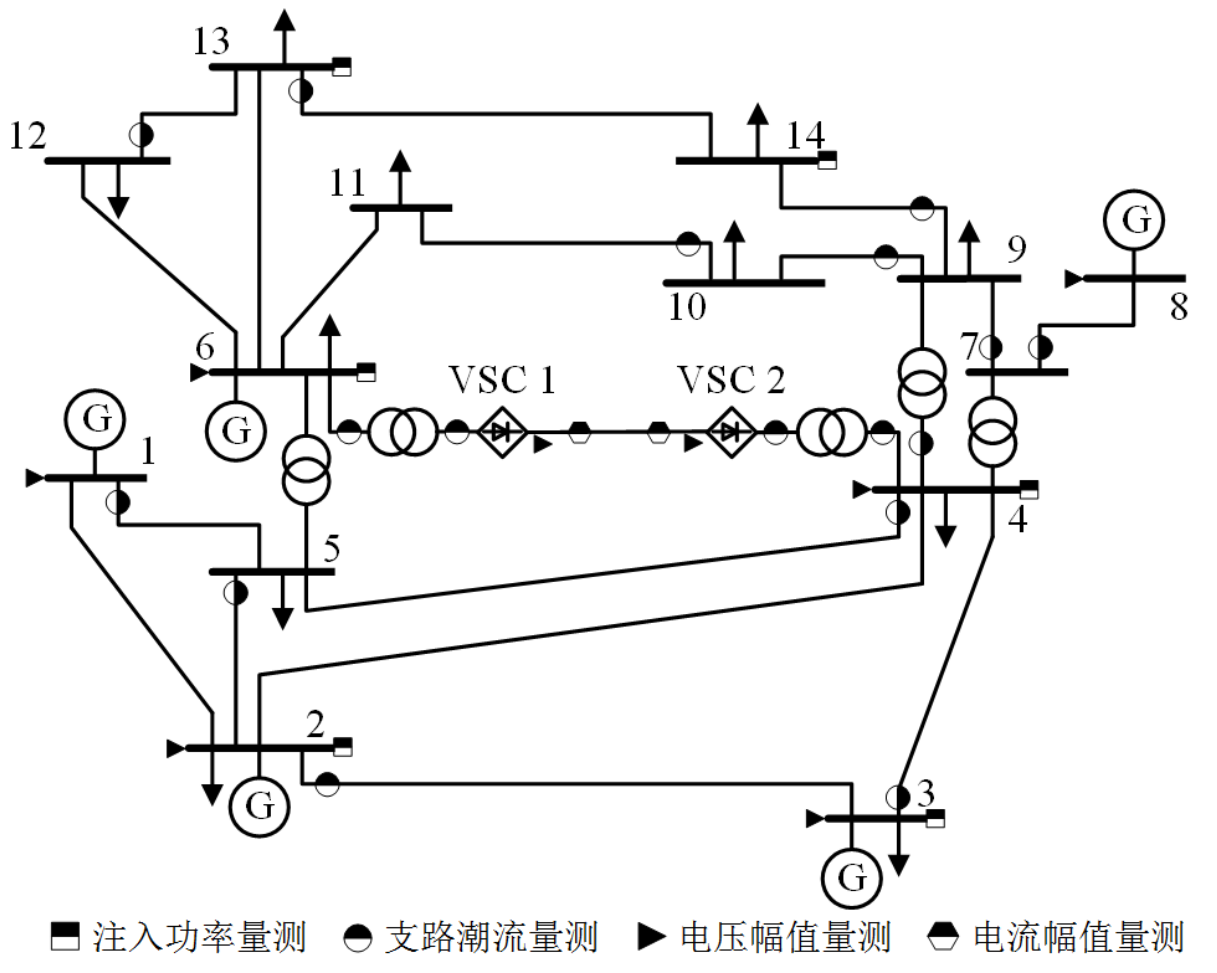

图 D2　第 2 组量测配置

**Fig. D2　Measurement configurations of group 2**

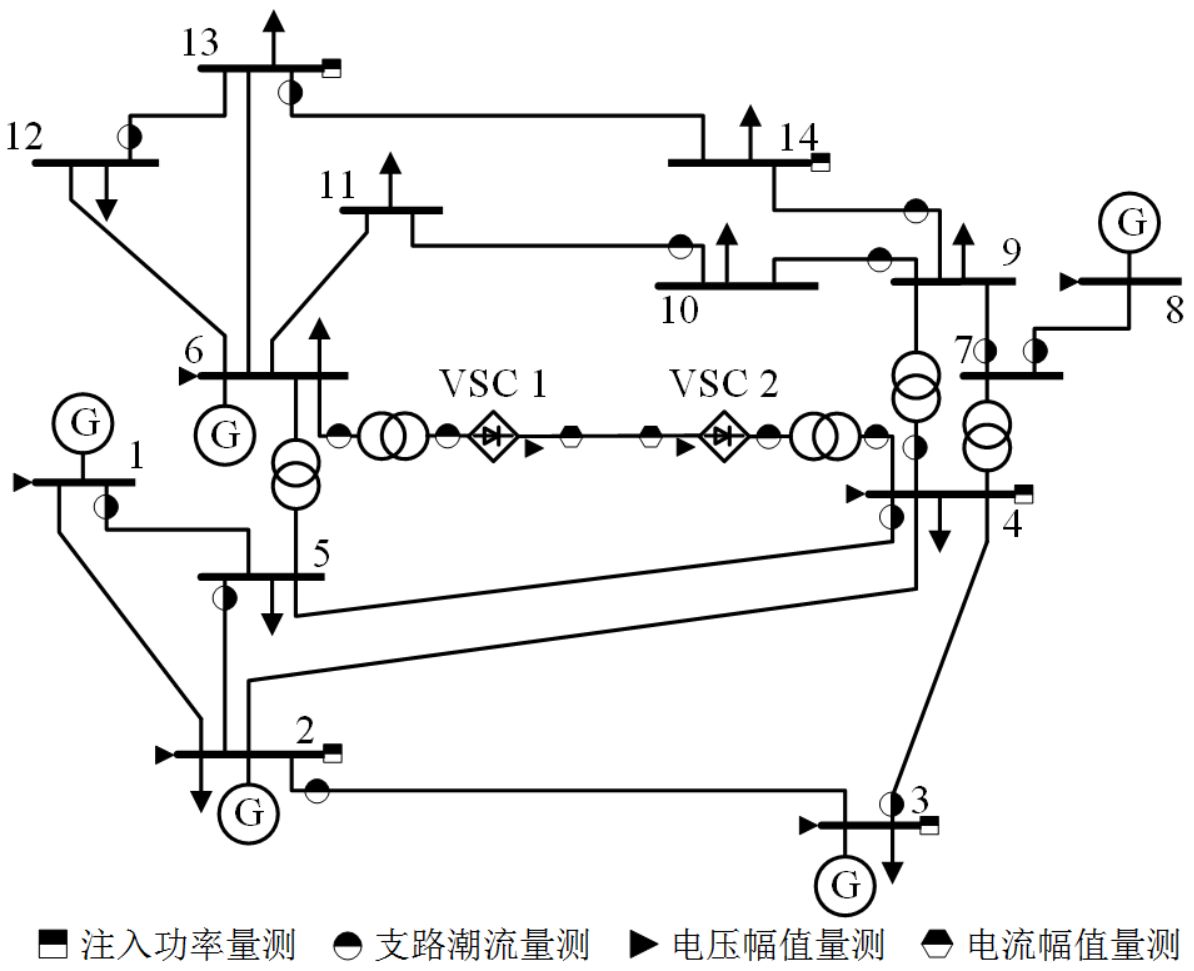

图 D3　第 3 组量测配置

**Fig. D3　Measurement configurations of group 3**

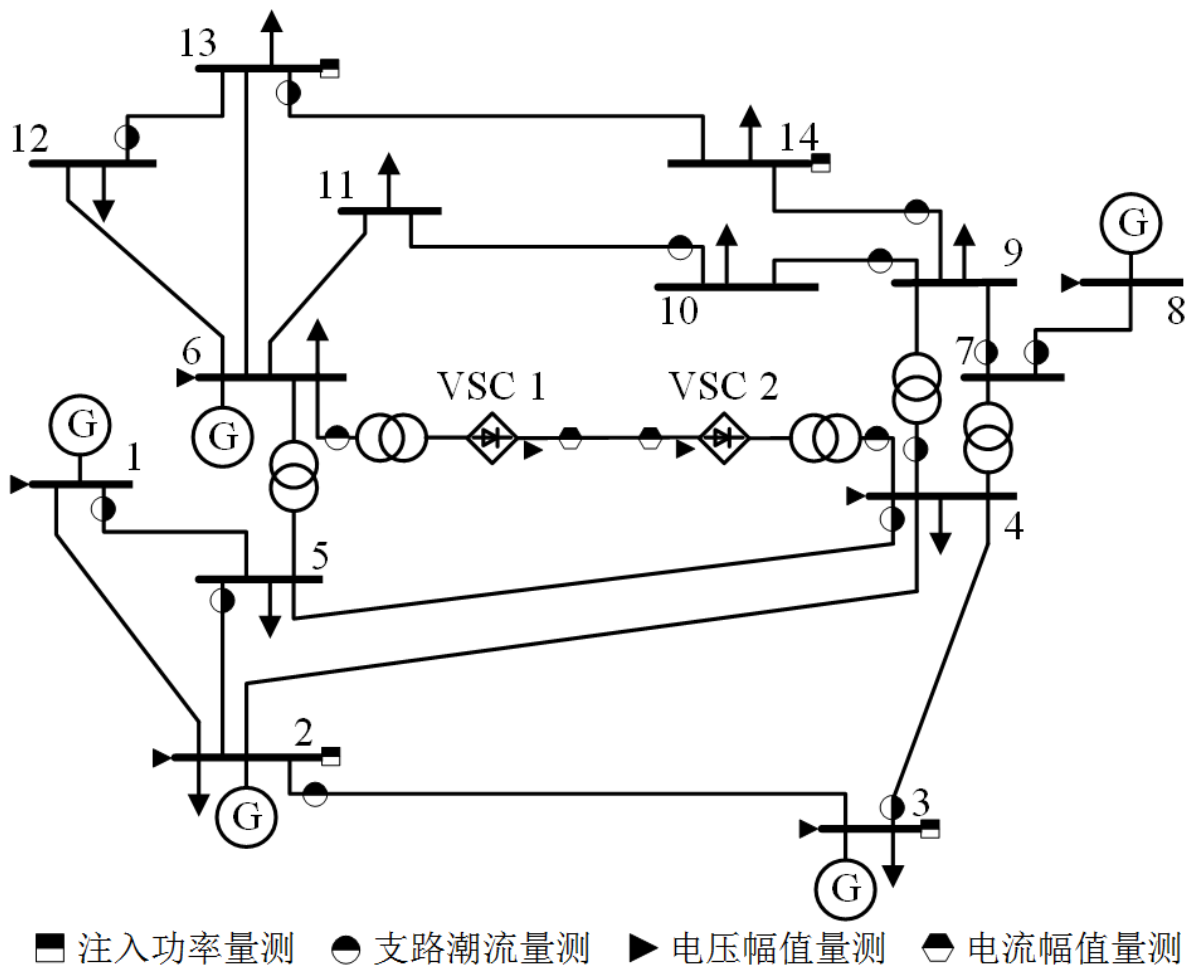

图 D6　第 6 组量测配置

**Fig. D6　Measurement configurations of group 6**

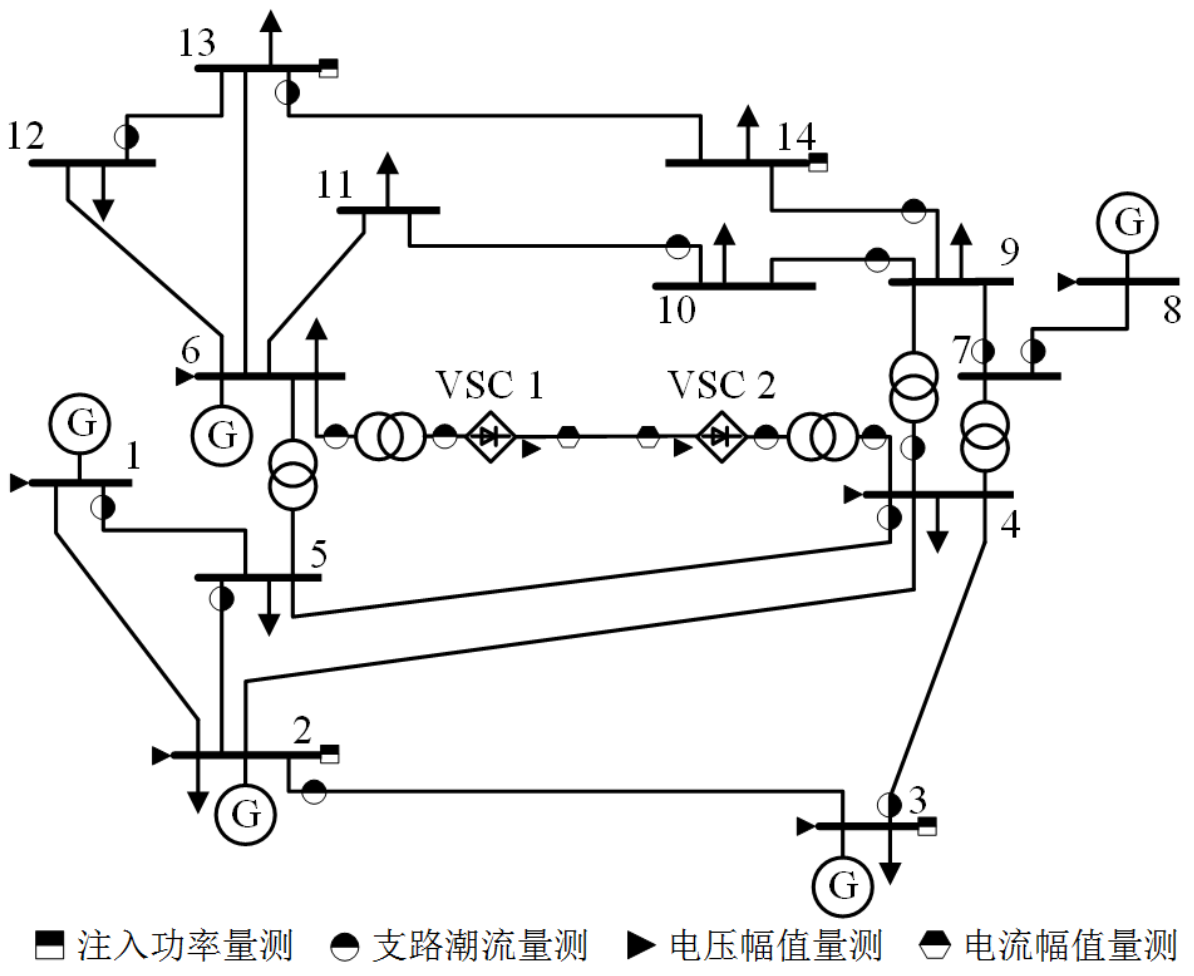

图 D4　第 4 组量测配置

**Fig. D4　Measurement configurations of group 4**

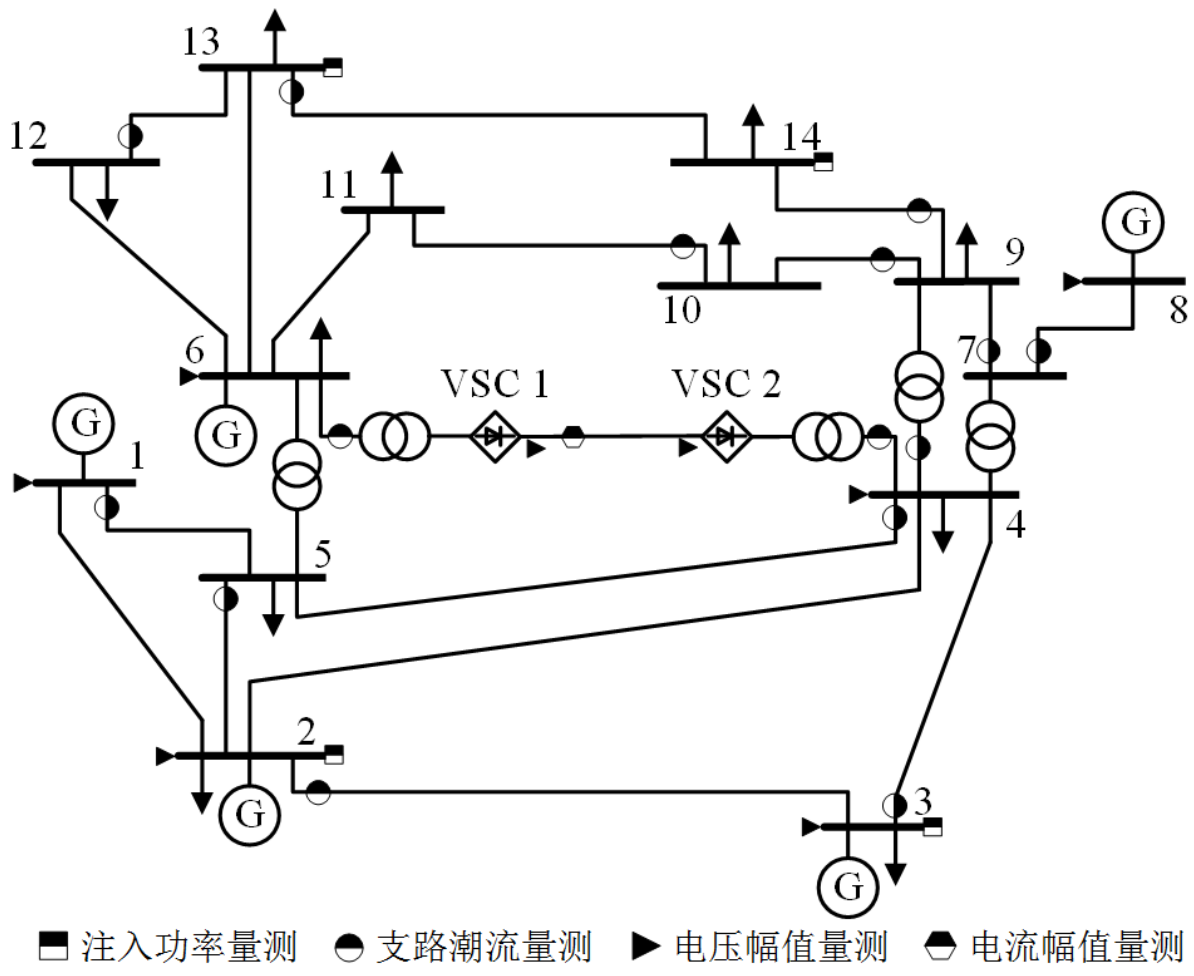

图 D7　第 7 组量测配置

**Fig. D7　Measurement configurations of group 7**

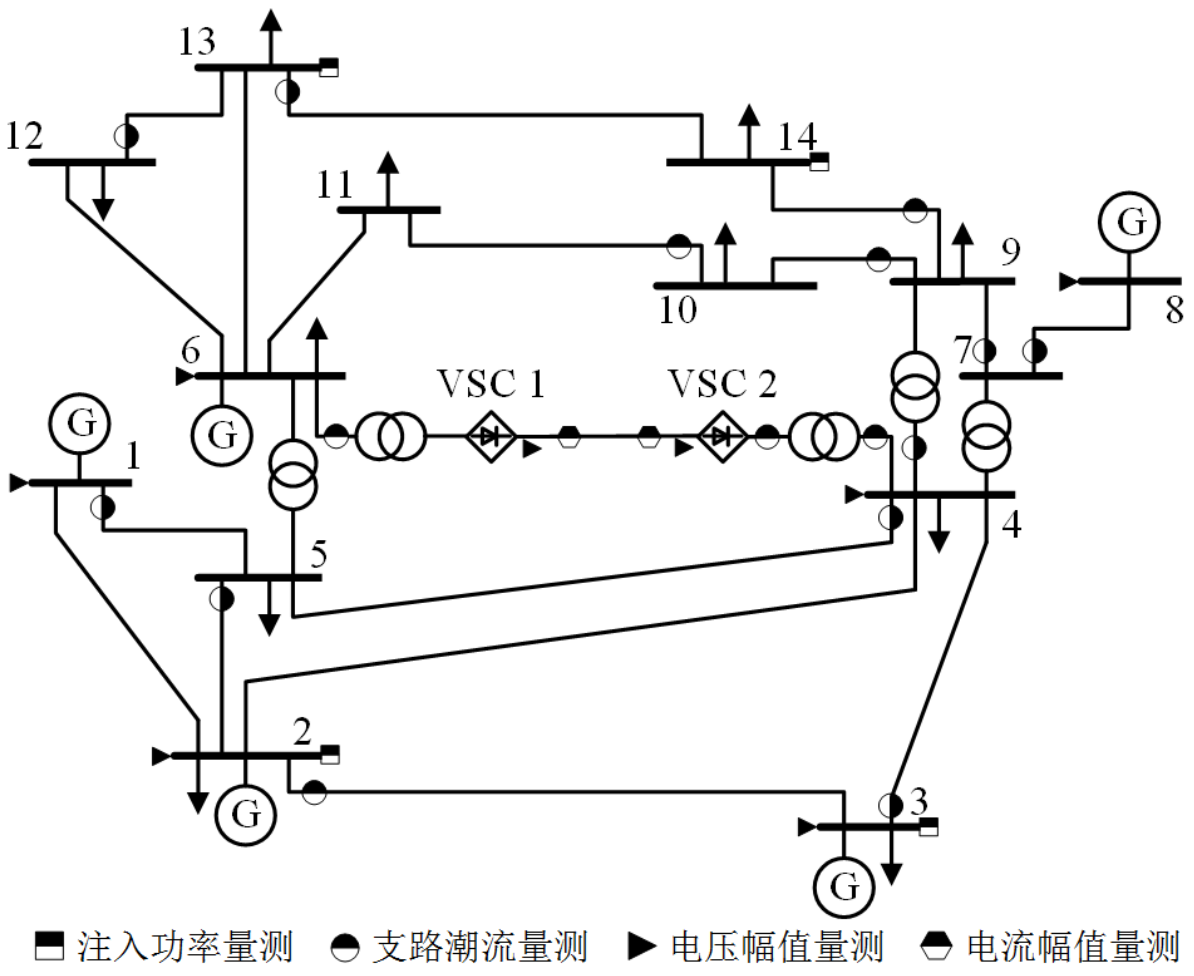

图 D5　第 5 组量测配置

**Fig. D5　Measurement configurations of group 5**

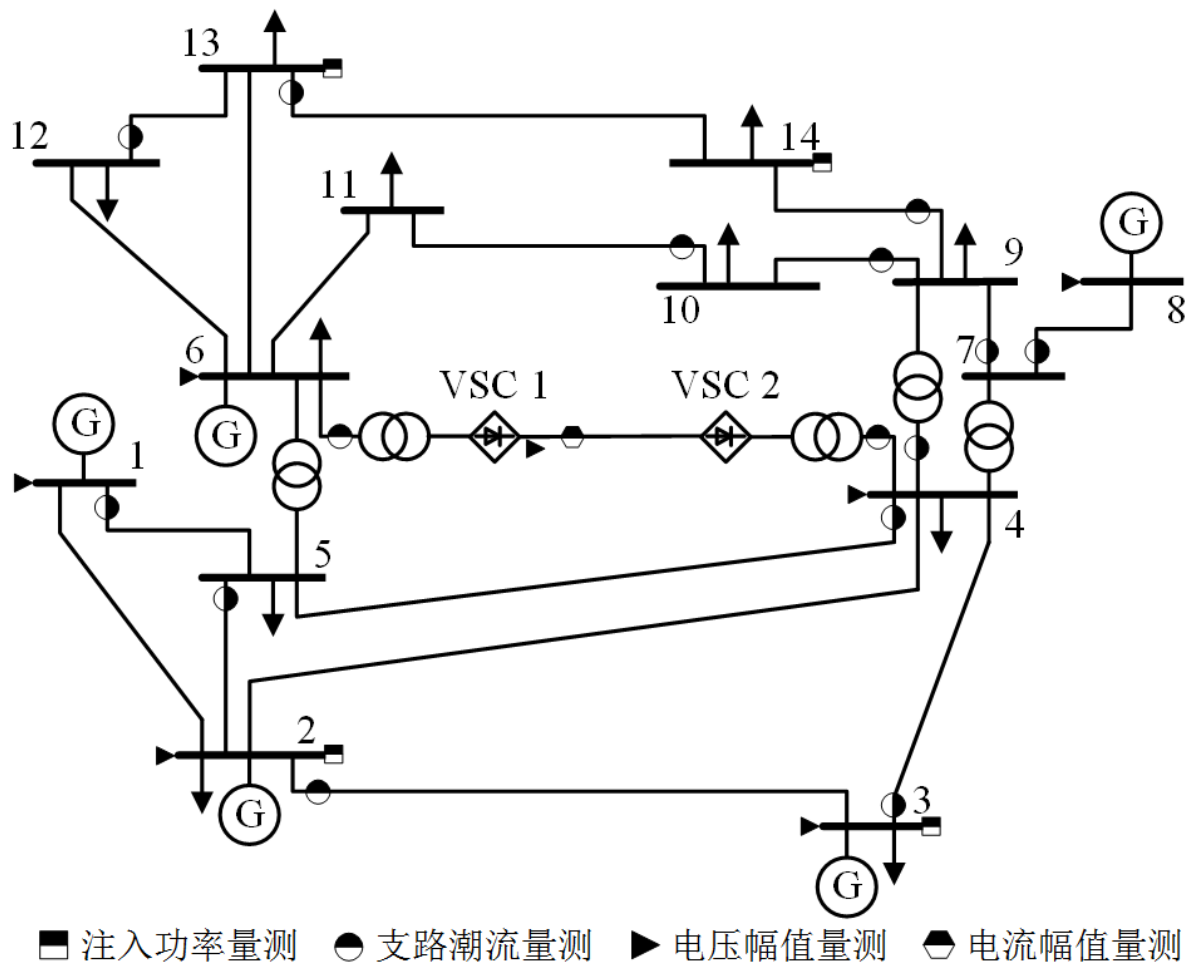

图 D8　第 8 组量测配置

**Fig. D8　Measurement configurations of group 8**

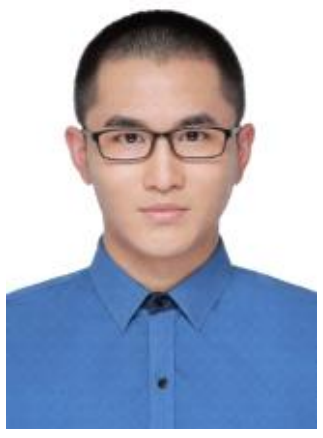

韩通

**作者简介**：

韩通(1993)，男，硕士研究生，研究方向为电力信息物理系统的恶意数据注入攻击问题，E-mail：hantong.eee@gmail.com；

陈艳波(1982)，男，通信作者，博士，副教授，硕导，主要研究方向为电力系统状态估计、信息安全、电力系统稳定与控制；

马进(1975)，男，博士，教授，博导，主要研究方向为电力系统稳定与控制。